\documentclass[aps,pra,showpacs,amssymb,nofootinbib,superscriptaddress,twocolumn,longbibliography]{revtex4-1}
\pdfoutput=1
\usepackage[utf8]{inputenc}
\usepackage[english]{babel}

\usepackage{tikz}
\usetikzlibrary{shapes,backgrounds,fit,decorations.pathreplacing,arrows,decorations.markings}
\usepackage{verbatim}
\tikzstyle{vecArrow} = [thick, decoration={markings,mark=at position
   1 with {\arrow[semithick]{open triangle 60}}},
   double distance=1.4pt, shorten >= 5.5pt,
   preaction = {decorate}
   postaction = {draw,line width=1.4pt, white,shorten >= 4.5pt}]
\tikzstyle{innerWhite} = [semithick, white,line width=1.4pt, shorten >= 4.5pt]

\usepackage{bbm}
\usepackage{bm}
\usepackage{amsbsy}
\usepackage{amsthm}
\usepackage{amssymb}
\usepackage{amsfonts}
\usepackage{amsmath}
\usepackage{dsfont} 
\usepackage{graphicx} 
\usepackage{epsfig}
\usepackage{epstopdf}
\usepackage{dsfont}
\usepackage{multibib}
\usepackage{color}

\usepackage[colorlinks]{hyperref}
\makeatletter
\newcommand\org@hypertarget{}
\let\org@hypertarget\hypertarget
\renewcommand\hypertarget[2]{%
  \Hy@raisedlink{\org@hypertarget{#1}{}}#2%
  }
\makeatother
\usepackage[figure,table]{hypcap}
\usepackage{MnSymbol}
\usepackage{enumerate}
\usepackage{float}
\hypersetup{
	bookmarksnumbered,
	pdfstartview={FitH},
	citecolor={darkgreen},
	linkcolor={darkred},
	urlcolor={darkblue},
	pdfpagemode={UseOutlines}}
\definecolor{darkgreen}{RGB}{50,190,50}
\definecolor{darkblue}{RGB}{0,0,190}
\definecolor{darkred}{RGB}{238,0,0}
\definecolor{quantum}{RGB}{83,37,127}
\definecolor{quantumlight}{RGB}{169,146,191}
\usepackage{soul}

\newcommand{\ket}[1]{\ensuremath{\left|\right.\!{#1}\!\left.\right\rangle}}

\newcommand{\bra}[1]{\ensuremath{\left\langle\right.\!{#1}\!\left.\right|}}

\newcommand{\ketbra}[2]{\ensuremath{|{#1}\rangle\!\langle{#2}|}}
\newcommand{\brakket}[3]{\ensuremath{\langle{#1}|{#2}|{#3}\rangle}}

\newcommand{\djj}{d\kern-0.4em\char"16\kern-0.1em}
\makeatletter
\renewcommand{\p@subsection}{}
\renewcommand{\p@subsubsection}{}
\makeatother

\usepackage{soul}

\usepackage[customcolors]{hf-tikz}
\tikzset{style green/.style={
    set fill color=green!50!lime!60,
    set border color=white,
  },
  style cyan/.style={
    set fill color=cyan!90!blue!60,
    set border color=white,
  },
  style orange/.style={
    set fill color=orange!80!red!60,
    set border color=white,
  },
  style hordash/.style={
    set fill color=white,
    set border color=black,
  },
     style rose/.style={
    set fill color= magenta!70!pink!70, 
    set border color=white,
  },
  hor/.style={
    above left offset={-0.09,0.25},
    below right offset={0.09,-0.05},
    #1
  },
  ver/.style={
    above left offset={-0.09,0.35},
    below right offset={0.09,-0.1},
    #1
  }
}

\usepackage[framemethod=tikz]{mdframed}
\definecolor{mycolor}{rgb}{0.122, 0.435, 0.698}
\newmdenv[innerlinewidth=0.5pt, roundcorner=4pt,linecolor=mycolor,innerleftmargin=6pt,
innerrightmargin=6pt,innertopmargin=6pt,innerbottommargin=6pt]{mybox}
\usepackage{paracol}
\usepackage{tcolorbox}
\tcbuselibrary{theorems}

\newtcbtheorem{Definitions}{Definition}%
{colback=mycolor!5,colframe=mycolor,fonttitle=\bfseries,
left=4pt,
right=4pt,
top=5pt,
bottom=5pt}{defi}

\newtcbtheorem{Lemmas}{Lemma}%
{colback=green!5,colframe=green!50!blue,fonttitle=\bfseries,
left=4pt,
right=4pt,
top=5pt,
bottom=5pt
}{lemma}
\newtcbtheorem{Results}{Result}%
{colback=orange!5,colframe=orange!50!blue,fonttitle=\bfseries,
left=4pt,
right=4pt,
top=5pt,
bottom=5pt,
title={#2},
}{result}
\newtcolorbox[blend into=figures]{boxdefi}[3][]
{ float*=ht,width=\textwidth,lower separated=false, center upper,
title={#2},label= def:#3,#1}

\date{\today}

\begin{document}

\title{Simplifying the design of multi-level thermal machines using virtual qubits}

\author{Ayaka Usui}
\email{ayaka.usui@icc.ub.edu}
\affiliation{Quantum Systems Unit, Okinawa Institute of Science and Technology Graduate University, Onna, Okinawa 904-0495, Japan}

\author{Wolfgang Niedenzu}
\email{Wolfgang.Niedenzu@uibk.ac.at}
\affiliation{Institut f\"{u}r Theoretische Physik, Universit\"{a}t Innsbruck, Technikerstra\ss e 21a, A-6020 Innsbruck, Austria}

\author{Marcus Huber}
\email{marcus.huber@univie.ac.at}
\affiliation{Institute for Quantum Optics and Quantum Information - IQOQI Vienna, Austrian Academy of Sciences, Boltzmanngasse 3, 1090 Vienna, Austria}
  \affiliation{Vienna Center for Quantum Science and Technology, Atominstitut, TU Wien,  1020 Vienna, Austria}%
\begin{abstract}
    Quantum thermodynamics often deals with the dynamics of small quantum machines interfacing with a large and complex environment. Virtual qubits, collisional models and reset master equations have become highly useful tools for predicting the qualitative behaviour of two-dimensional target systems coupled to few-qubit machines and a thermal environment. 
    While few successes in matching the simplified model parameters for all possible physical systems are known, the qualitative predictions still allow for a general design of quantum machines irrespective of the implementation.
    We generalise these tools by introducing multiple competing virtual qubits for modelling multi-dimensional systems coupled to larger and more complex machines. By simulating the full physical dynamics for targets with three dimensions, we uncover general properties of reset models that can be used as `dials' to correctly predict the qualitative features of physical changes in a realistic setup and thus design autonomous quantum machines beyond a few qubits. We then present a general analytic solution of the reset model for arbitrary-dimensional systems coupled to multi-qubit machines. Finally, we showcase an improved three-level laser as an exemplary application of our results.
\end{abstract}
\maketitle

\section{Introduction}

Machines operating at the quantum scale offer an exploration of the ultimate limits of thermodynamic tasks \cite{alicki1979quantum,kosloff1984quantum,kosloff2013quantum,gelbwaser2015thermodynamics,goold2016role,vinjanampathy2016quantum,ghosh2019thermodynamic,bhattacharjee2020quantumthermal}, such as cooling down individual quantum systems or creating coherent sources of light. Design and control of such processes is usually assumed and achieved at the level of few quantum mechanical degrees of freedom, interacting with a large environment that one lacks detailed control over \cite{koski2014experimental,rossnagel2016single,klaers2017squeezed,peterson2019experimental,vonlindenfels2019spin,klatzow2019experimental}. As large quantum systems are notoriously hard to simulate exactly, most of the focus is devoted to deriving master equations and dynamics for few-qubit machines or a single qutrit interacting with multiple baths. A crucial discovery in that context is the concept of a virtual qubit~\cite{brunner2012virtual,Linden2010,skrzypczyk2011smallest}. 
It allows one to extract an effective two-level transition from multiple levels by sacrificing detailed knowledge of the machine behaviours and focusing only on the system of interest. Therefore, one can dramatically reduce the complexity of predicting the steady state and even transient dynamics~\cite{Silva2016,Erker2017,Seah2018,Manzano2019,man2017smallest,chen2017thermodynamic,clivaz2019unifying}.
Going beyond simple qubit targets, however, is a challenge due to the potential complexity of competing interactions with multiple virtual qubits. 

In this paper, we solve the problem, for arbitrary-dimensional quantum systems (qudits), interacting with multiple competing virtual qubits across all possible two-level transitions in the context of reset-type master equations. We explore the solution for three-level quantum systems and compare it to optical master equations, identifying a few universal features that these approaches share and thus important properties of complex machine designs that this simple and computable model correctly predicts. Finally, we use our model and analysis to study an enhancement of the paradigmatic three-level maser/laser~\cite{scovil1959three} through more complex machines. 

We note that our approach simplifies the complexity of finding the steady state if the virtual qubits are characterised, but is not applicable for the transient regime unless the natural description of the system dynamics is done in terms of virtual qubits in the first place. Nevertheless, the performance of autonomous machines, is mostly encoded in the steady state. For instance, an autonomous refrigerator has some transient behaviour depending on the initial state but then approaches a non-equilibrium steady state, with a constant transport of heat away from a target towards the environment~{\cite{Mitchison2015}}. It is exactly this steady state that encodes the final temperature stabilised within a target system. More importantly, the simplification provided by virtual qubits depends on what parameters about complex systems are known and what the most appropriate method for modelling them is. 
By utilising the results obtained in this work, we show in the strong dissipation limit that (i) given a qudit system coupled to some two-qubit machines, the steady state of the qudit can be derived and that (ii) given a desired steady state of the qudit, the parameters in the machines can be tuned.


\section{Motivation: The two-qubit machine as a virtual qubit}

First, we review the idea of virtual qubits, which has been proposed in Ref.~\cite{brunner2012virtual}. Consider two qubits with energy spacings $\Omega_1$ and $\Omega_2$ (we assume $\Omega_1>\Omega_2$) that coherently interact with each other and are in contact with two thermal baths at temperatures $T_1$ and $T_2$, respectively. This two-qubit machine is composed of the energy eigenstates in the absence of coherent coupling, $\ket{0}_{\mathrm{1}}\ket{0}_{\mathrm{2}}$, $\ket{0}_{\mathrm{1}}\ket{1}_{\mathrm{2}}$, $\ket{1}_{\mathrm{1}}\ket{0}_{\mathrm{2}}$, and $\ket{1}_{\mathrm{1}}\ket{1}_{\mathrm{2}}$. The single-excitation manifold is then called a \emph{virtual qubit} whose ground and excited state are given by $\ket{0}_{\mathrm{1}}\ket{1}_{\mathrm{2}}$ and $\ket{1}_{\mathrm{1}}\ket{0}_{\mathrm{2}}$, respectively, with the energy spacing $\Omega_1 - \Omega_2$. The temperature of this virtual qubit, called the virtual temperature, is determined by the ratio between the ground and excited state populations, which, together with the Boltzmann law, leads to
\begin{equation} \label{eq:Tv}
    T_{\mathrm{v}}
    =
    \frac
    {\Omega_1 - \Omega_2}
    {\Omega_1/T_1 - \Omega_2/T_2}
    \;.
\end{equation}
Note that, since it is not a real temperature, $T_\mathrm{v}$ may be negative in the case of population inversion. 
The levels of the virtual qubit discussed here are $\ket{0}_{1}\ket{1}_{2}$ and $\ket{1}_{1}\ket{0}_{2}$. We note, that what matters is only the population bias, therefore, in principle, any level structure would work.

\begin{figure}
    \centering
    \includegraphics[width=0.95\columnwidth]{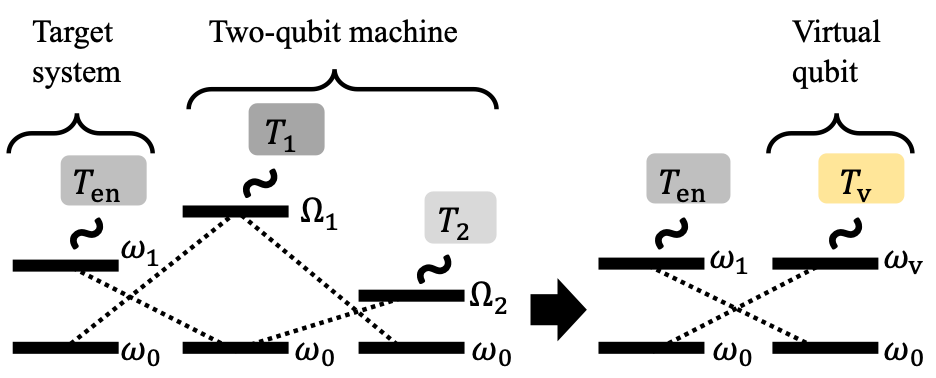}
    \caption{Sketch of a virtual qubit. Left: a target qubit coupled to a two-qubit machine, where $\omega_1-\omega_0=\Omega_1-\Omega_2$. Right: the target qubit effectively coupled to a virtual qubit, where $\omega_{\mathrm{v}}-\omega_0=\Omega_1-\Omega_2$ and $T_{\mathrm{v}}$ is given by Eq.~\eqref{eq:Tv}. The dotted lines represent the coherent interactions given in Eq.~\eqref{eq:H_3qubit}. The wavy lines represent contact with baths whose temperatures are $T_{\mathrm{en},1,2}$.
    } \label{fig:qubit_vq_Ten}
\end{figure}

We now add another physical qubit, namely the \emph{target qubit}, with energy spacing $\omega_1-\omega_0=\Omega_1-\Omega_2$ that is coherently coupled to the two-qubit machine (see Fig.~\ref{fig:qubit_vq_Ten}). Assuming that this target qubit is further interacting with an environment at temperature $T_{\mathrm{en}}$, 
the dynamics of the composite system are determined by the reset master equation (RME)
(see Refs~\cite{Linden2010,skrzypczyk2011smallest} or Appendix~{\ref{app:RME}} for the derivation of this RME)
\begin{align}\label{eq:master_1vq_en}
    \frac{\partial \rho_{\mathrm{tot}}}{\partial t}
    =
    -i\left[H,\rho_{\mathrm{tot}}\right]
    &+
    Q_{\mathrm{en}} \left(
    \tau_{\mathrm{en}}\otimes \mathrm{Tr}_{\mathrm{tar}}[\rho_{\mathrm{tot}}] - \rho_{\mathrm{tot}}
    \right)
    \nonumber\\
    &
    +
    Q_{1} \left(
    \tau_{1}\otimes\mathrm{Tr}_{1}[\rho_{\mathrm{tot}}] - \rho_{\mathrm{tot}}
    \right)
    \nonumber\\
    &
    +
    Q_{2} \left(
    \tau_{2}\otimes\mathrm{Tr}_{2}[\rho_{\mathrm{tot}}] - \rho_{\mathrm{tot}}
    \right)
    ,
\end{align} 
where $\rho_{\mathrm{tot}}$ is the density matrix of the composite system. Here, $Q_{\mathrm{en},1,2}$ are thermalisation rates corresponding to the environment or the thermal baths in contact with the two-qubit machine, respectively. The density matrices $\tau_{\mathrm{en},1,2}$ are thermal states corresponding to the real temperatures $T_{\mathrm{en},1,2}$, respectively. The partial traces over the target qubit or the machine's constituents are denoted $\mathrm{Tr}_{\mathrm{tar},1,2}$, respectively. The Hamiltonian in Eq.~\eqref{eq:master_1vq_en} reads
\begin{equation}\label{eq:H_3qubit}
    H
    =
    \sum_{k=0}^{1}
    \omega_k \ketbra{k}{k}
    +
    \sum_{i\in\{1,2\}}\!\!
    \Omega_{i}
    \sigma_{i}^{+}\sigma_{i}^{-}
    +
    g 
    \ketbra{0}{1}\sigma_{1}^{+}\sigma_{2}^{-}
    +
    \mathrm{H.c.}
\end{equation}
with $\sigma^+_i=\ket{1}_i\bra{0}_i$ and the coherent coupling strength $g$. 
We assume that this coupling is sufficiently weak for the system to be identified as an isolated entity so that any transition between two non-degenerate levels of the system is accompanied by a corresponding exchange in the machine and the environment.
Without the environment, the dynamics drive the target qubit into a steady state at the virtual temperature~\eqref{eq:Tv}, independent of the rates $Q_1$ and $Q_2$ \cite{brunner2012virtual}.

In general, however, the two-qubit machine is disturbed by the target qubit's interaction with the environment and hence the virtual temperature \eqref{eq:Tv} is not the steady-state temperature anymore \cite{skrzypczyk2011smallest}. However, if the qubits inside the two-qubit machine thermalise very fast with the two baths at temperatures $T_{1,2}$, i.e., if $Q_{1,2}\gg Q_{\mathrm{en}},g$, the notion of the virtual temperature~\eqref{eq:Tv} remains valid. 

Here, assuming $Q_{1,2}\gg Q_{\mathrm{en}},g$, we replace the two-qubit machine with a bath at the virtual temperature \eqref{eq:Tv} and consider the effective reset master equation (effRME) for the target system only,
\begin{equation} \label{eq:master_1vq_en_simplified}
    \frac{\partial \rho}{\partial t}
    =
    Q_{\mathrm{en}} \left(
    \tau_{\mathrm{en}}
    - 
    \rho
    \right)
    +
    q_{\mathrm{vir}} \left(
    \tau_{\mathrm{vir}}
    - 
    \rho
    \right)
    ,
\end{equation}
where $q_{\mathrm{vir}}$ is the effective thermalisation rate to the virtual qubit and $\tau_{\mathrm{vir}}$ is a thermal state at the virtual temperature \eqref{eq:Tv}. Note that we focus on the steady-state regime, and therefore in this model all off-diagonal terms in the density matrix vanish. This present approach does not work if one is interested in, for example, the transient time~\cite{Brask2015,Mitchison2015} or a different model having coherent transitions, both of which coherence plays a role in. Furthermore, note that the target Hamiltonian commutes with the dissipator of the effRME, and this approximation is valid only for this class of Hamiltonians.

The steady-state solution of this effRME reads
\begin{equation} \label{eq:steady_1vq_en}
    \rho_{\mathrm{ss}}
    =
    C
    \left(
    Q_{\mathrm{en}} \tau_{\mathrm{en}}
    +
    q_{\mathrm{vir}} \tau_{\mathrm{vir}}
    \right)
\end{equation}
with the normalisation $C=(Q_{\mathrm{en}}+q_{\mathrm{vir}})^{-1}$. Note that owing to the two competitive dissipative couplings, this steady-state solution explicitly depends on the rates $Q_\mathrm{en}$ and $q_\mathrm{vir}$.

Comparing this steady state \eqref{eq:steady_1vq_en} of the effRME with that from the RME~\eqref{eq:master_1vq_en}, 
and using $Q_{1,2}\gg Q_{\mathrm{en}},g$, we find
\begin{equation} \label{eq:qvir}
    q_{\mathrm{vir}}
    =
    \frac{2g^2}{Q_1 + Q_2}
    \left(
    \tau_{1}^{\mathrm{g}} \tau_{2}^{\mathrm{e}}
    +
    \tau_{1}^{\mathrm{e}} \tau_{2}^{\mathrm{g}}
    \right).
\end{equation}
Here, $\tau_{1,2}^{\mathrm{g,e}}$ are the populations of the ground and excited states of the thermal state at temperatures $T_{1,2}$, respectively. 
We denote the \emph{norm} of the virtual qubit
\begin{align} \label{eq:norm_vir}
    n_{\mathrm{vir}}
    &=
    \tau_{1}^{\mathrm{g}}\tau_{2}^{\mathrm{e}}+\tau_{1}^{\mathrm{e}}\tau_{2}^{\mathrm{g}}
\end{align}
as it corresponds to the weight of the levels $\ket{0}_1 \ket{1}_2$ and $\ket{1}_1 \ket{0}_2$ that form the virtual qubit within the two-qubit machine space. This norm thus determines the temperature dependence of the effective rate $q_\mathrm{vir}$. 

\section{Three-level system coupled to three two-level machines}

\begin{figure}
    \centering
    \includegraphics[width=0.95\columnwidth]{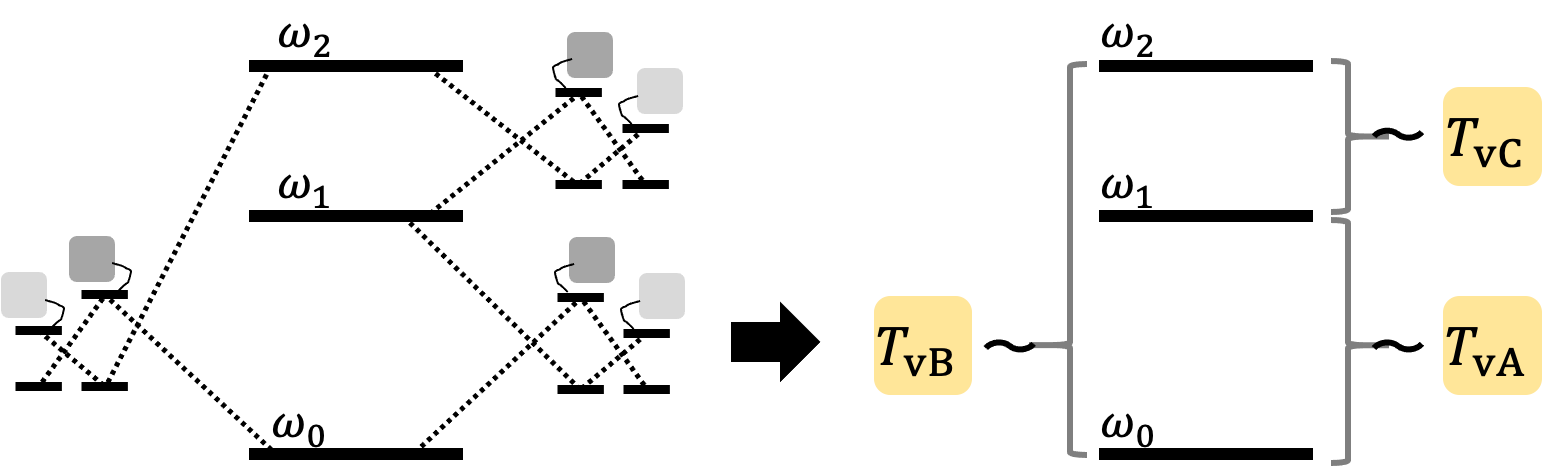}
    \caption{
    Simplification of two-qubit machines by using virtual qubits. Left: a qutrit coupled to three two-qubit machines. The dotted lines represent coherent interactions. Right: the qutrit where all the two-qubit machines are assumed to be baths at their virtual temperatures.} \label{fig:qutrit_6qubits}
\end{figure}

We now continue by applying the idea of virtual qubits to higher-dimensional target systems. In this section, we consider a three-level system, i.e., a qutrit, with energy spacings $\omega_{0,1,2}$ that is coupled to several two-qubit machines (Fig.~\ref{fig:qutrit_6qubits}). Within the effRME description, each of these two-qubit machines is regarded as a virtual qubit.

With a single machine coupled to the target qutrit, the situation is essentially the same as the qubit target shown Fig.~\ref{fig:qubit_vq_Ten}. With two machines coupled, two distinct thermalisation processes act on the target. Consequently, the steady state of the latter is not a Gibbs-like state, unless both virtual temperatures are the same.
As an example, if a machine with virtual temperature $T_{\mathrm{v}1}$ is connected to the target levels $\ket{0}$ and $\ket{1}$ and another one with virtual temperature $T_{\mathrm{v}2}$ is connected to the levels of $\ket{0}$ and $\ket{2}$, the qutrit is driven into the steady state 
 \begin{equation}
 \rho_{\mathrm{ss}}=C \left( \ketbra{0}{0} + \mathrm{e}^{-(\omega_1-\omega_0)/T_{\mathrm{v}1}} \ketbra{1}{1} + \mathrm{e}^{-(\omega_2-\omega_0)/T_{\mathrm{v}2}} \ketbra{2}{2} \right)
 \end{equation}
 with the normalisation $C=(1+\mathrm{e}^{-(\omega_1-\omega_0)/T_{\mathrm{v}1}} + \mathrm{e}^{-(\omega_2-\omega_0)/T_{\mathrm{v}2}})^{-1}$. As the two thermalisation processes do not compete, each transition is ``thermalised'' to its respective virtual temperature $T_{\mathrm{v}1},T_{\mathrm{v}2}$. 
By contrast, if all transitions within the target qutrit interact with independent two-level machines (see Fig.~\ref{fig:qutrit_6qubits}), the three thermalisation process compete against each other unless all the virtual temperatures are equal. 

Below, we utilise the idea of the virtual qubits to construct an effRME of the three-level target system as we did for the qubit target system. To explore the parameter dependency of the effective thermalisation rates, we compare the steady state among the effRME for the target system and two non-exclusive physical models for the corresponding machine setup: the full RME for target plus machine and an optical master equation, i.e., a so-called Gorini-Kossakowski-Lindblad-Sudarshan master equation (GKLSME). Finally, we discuss these models' relations to the effRME. 

Although we stick to the qutrit target system throughout this section, we have considered $n$-dimensional target systems in Appendix~{\ref{app:derivation_steady_n}} and solved the effRME for the steady state by using the treatment introduced in this section. Additionally, we present the steady-state solution for a four-dimensional target system in Appendix~{\ref{app:steady_4}}.

\subsection{Effective reset master equation (effRME)}\label{sec_qutrit}

\begin{figure}
    \centering
    \includegraphics[width=0.98\columnwidth]{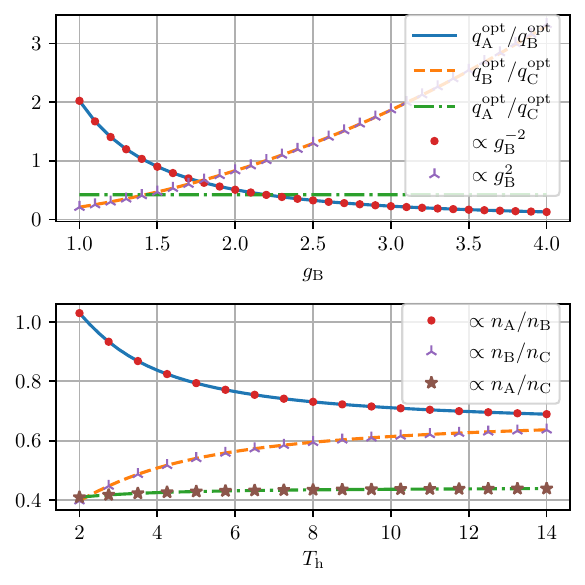}
    \caption{
    Optimal coupling coefficient rates between the target qutrit and the auxiliary qubits for which the state~\eqref{eq:steady_3} is the steady-state solution of the RME~\eqref{eq:master_reset} as a function of coherent interaction strength $g_{\mathrm{B}}$ (upper panel) and bath temperature $T_{i1}=T_{\mathrm{h}}$ (lower panel). 
    In the upper panel, the ratio $q_{\mathrm{A}}/q_{\mathrm{B}}$ is proportional to $1/g_{\mathrm{B}}^2$, and the ratio $q_{\mathrm{B}}/q_{\mathrm{C}}$ is proportional to $g_{\mathrm{B}}^2$. 
    In the lower panel, the ratios $q_i/q_j$ are proportional to the ratios $n_i/n_j$ of the norms~\eqref{eq_norm} of the virtual qubits for $i,j\in\{\mathrm{A,B,C}\}$.
    Both plots use the same parameter set, except for $T_{i1}=T_{\mathrm{h}}=3.1$ in the upper panel and $g_{\mathrm{B}}=1.5$ in the lower panel.
    As mentioned in the text, we take $\omega_0=0$ and set the energy unit as $(\omega_1-\omega_0)/2$. Thus, accordingly, $\omega_1=2$. As for other parameters, the following is used:
    $\omega_2=3$, $\Omega_{\mathrm{A}1}=2.5$, $\Omega_{\mathrm{B}1}=4.5$, $\Omega_{\mathrm{C}1}=1.3$, $g_\mathrm{A}=1.2$, $g_\mathrm{C}=1.8$, $T_{i1}=T_{\mathrm{h}}=3.1$, $T_{i2}=T_{\mathrm{c}}=1.2$, $Q_{i1}=70$, and $Q_{i2}=50$ for $i\in\{\mathrm{A},\mathrm{B},\mathrm{C}\}$.
    }
    \label{fig:ratioq_reset}
\end{figure}

We consider a qutrit coherently coupled to three pairs of two physical qubits, as depicted in Fig.~\ref{fig:qutrit_6qubits}. We label as ``A'' the pair coupled to the levels of $\ket{0}$ and $\ket{1}$, as ``B'' the pair coupled to the levels of $\ket{0}$ and $\ket{2}$, and as ``C'' the pair coupled to the levels of $\ket{1}$ and $\ket{2}$. Each of the pairs has two qubits 
with energy spacings $\Omega_{i1}$ and $\Omega_{i2}$, 
and the qubits are in contact baths, the temperatures of which are $T_{i1}$ and $T_{i2}$, respectively, for $i\in\{\mathrm{A},\mathrm{B},\mathrm{C}\}$. Furthermore, due to energy conservation, the energy spacings are restricted as $\omega_{1}-\omega_{0}=\Omega_{\mathrm{A}1}-\Omega_{\mathrm{A}2}$, $\omega_{2}-\omega_{0}=\Omega_{\mathrm{B}1}-\Omega_{\mathrm{B}2}$, and $\omega_{2}-\omega_{1}=\Omega_{\mathrm{C}1}-\Omega_{\mathrm{C}2}$.

We assume that the thermalisation of the qubits inside the two-qubit machines is fast enough that the concept of the virtual temperature is valid. 
By considering that the two-qubit machines maintain their virtual temperatures, the effRME of the target system is provided by
\begin{equation}  \label{eq:master_ABC}
\begin{split}
    \frac{\partial \rho}{\partial t}
    &=
    \sum_{i\in\{\mathrm{A},\mathrm{B},\mathrm{C}\}}
    q_{i} \left(
    \tau_{i} \otimes \mathrm{Tr}_{i} [\rho] - \rho
    \right)
    ,
\end{split}
\end{equation}
where $q_{\mathrm{A,B,C}}$ are the effective thermalisation rates and $\mathrm{Tr}_{\mathrm{A,B,C}}$ represent tracing out the space of the qubit pairs A, B, C, respectively. The states $\tau_{\mathrm{A,B,C}}$ are thermal states at the virtual temperatures $T_{\mathrm{vA,vB,vC}}$, respectively, given by 
\begin{align} \label{eq:Tvj}
    T_{\mathrm{v}i}
    &=
    \frac{\Omega_{i1}-\Omega_{i2}}{\Omega_{i1}/T_{i1}-\Omega_{i2}/T_{i2}}
\end{align}
for $i\in\{\mathrm{A},\mathrm{B},\mathrm{C}\}$. Explicitly, these states read  $\tau_{\mathrm{A}}=\tau_{\mathrm{A}}^{\mathrm{g}} \ketbra{0}{0} + \tau_{\mathrm{A}}^{\mathrm{e}} \ketbra{1}{1}$, $\tau_{\mathrm{B}}=\tau_{\mathrm{B}}^{\mathrm{g}} \ketbra{0}{0} + \tau_{\mathrm{B}}^{\mathrm{e}} \ketbra{2}{2}$, and $\tau_{\mathrm{C}}=\tau_{\mathrm{C}}^{\mathrm{g}} \ketbra{1}{1} + \tau_{\mathrm{C}}^{\mathrm{e}} \ketbra{2}{2}$, respectively, where $\tau_{i}^{\mathrm{g,e}}$ are the respective populations of the ground and excited states. 
Each term in the summation of the effRME~{\eqref{eq:master_ABC}} describes thermalisation, and particularly $\tau_{i} \otimes \mathrm{Tr}_{i} [\rho]$ for $i\in\{\mathrm{A},\mathrm{B},\mathrm{C}\}$ means a state where the population ratio of the levels labelled by $i$ is $\exp[-\omega_i/T_{\mathrm{v}i}]$ with $\omega_{\mathrm{A}}=\omega_1-\omega_0$, $\omega_{\mathrm{B}}=\omega_2-\omega_0$, and $\omega_{\mathrm{C}}=\omega_2-\omega_1$. See Appendix~{\ref{app:partialtrace}} for details of how to calculate the partial traces $\mathrm{Tr}_{\mathrm{A,B,C}}[\rho]$.
By solving the effRME~\eqref{eq:master_ABC} for $\partial\rho/\partial t=0$, the steady state of the target system can be found as
\begin{equation} \label{eq:steady_3}
    \rho_{\mathrm{ss}}
    =
    C\left(
    q_{\mathrm{A}}
    q_{\mathrm{B}}
    \tau_{\mathrm{AB}}
    +
    q_{\mathrm{B}}
    q_{\mathrm{C}}
    \tau_{\mathrm{BC}}
    +
    q_{\mathrm{C}}
    q_{\mathrm{A}}
    \tau_{\mathrm{CA}}
    \right),
\end{equation}
where the normalisation is $C=(q_{\mathrm{A}}q_{\mathrm{B}}\mathrm{Tr}[\tau_{\mathrm{AB}}]+q_{\mathrm{B}}q_{\mathrm{C}}\mathrm{Tr}[\tau_{\mathrm{BC}}]+q_{\mathrm{C}}q_{\mathrm{A}}\mathrm{Tr}[\tau_{\mathrm{CA}}])^{-1}$. The steady state \eqref{eq:steady_3} is thus a combination of the respective steady states if only two of the three coherent couplings are present, i.e., 
\begin{subequations} \label{eq:tauABBCCA}
\begin{align}
    \tau_{\mathrm{AB}}
    &=
    \tau_{\mathrm{A}}^{\mathrm{g}}\tau_{\mathrm{B}}^{\mathrm{g}}
    \ketbra{0}{0}
    +
    \tau_{\mathrm{A}}^{\mathrm{e}}\tau_{\mathrm{B}}^{\mathrm{g}}
    \ketbra{1}{1}
    +
    \tau_{\mathrm{A}}^{\mathrm{g}}\tau_{\mathrm{B}}^{\mathrm{e}}
    \ketbra{2}{2},\\
    \tau_{\mathrm{BC}}
    &=
    \tau_{\mathrm{B}}^{\mathrm{g}}\tau_{\mathrm{C}}^{\mathrm{e}}
    \ketbra{0}{0}
    +
    \tau_{\mathrm{B}}^{\mathrm{e}}\tau_{\mathrm{C}}^{\mathrm{g}}
    \ketbra{1}{1}
    +
    \tau_{\mathrm{B}}^{\mathrm{e}}\tau_{\mathrm{C}}^{\mathrm{e}}
    \ketbra{2}{2},\\
    \tau_{\mathrm{CA}}
    &=
    \tau_{\mathrm{C}}^{\mathrm{g}}\tau_{\mathrm{A}}^{\mathrm{g}}
    \ketbra{0}{0}
    +
    \tau_{\mathrm{C}}^{\mathrm{g}}\tau_{\mathrm{A}}^{\mathrm{e}}
    \ketbra{1}{1}
    +
    \tau_{\mathrm{C}}^{\mathrm{e}}\tau_{\mathrm{A}}^{\mathrm{e}}
    \ketbra{2}{2}.
\end{align}
\end{subequations}
Note that these states are not normalised on purpose, i.e., $\mathrm{Tr}[\tau_{\mathrm{AB}}]\neq 1$, $\mathrm{Tr}[\tau_{\mathrm{BC}}]\neq 1$, and $\mathrm{Tr}[\tau_{\mathrm{CA}}]\neq 1$.


Although we are interested in the target qutrit in this section, we have generalised this virtual-qubit treatment to $n$-dimensional target systems in Appendix~{\ref{app:derivation_steady_n}}. A primary issue is that $n$-dimensional target systems possess $n(n-1)/2$ level pairs that can be coupled to more two-qubit machines, which increases the complexity of finding the steady state. However, by taking into account that the non-diagonal terms in the steady-state density matrix vanish, as seen in the effRMEs~{\eqref{eq:master_ABC}} or {\eqref{eq:master_kl}}, solving the effRME of a $n$-dimensional target system for $\partial \rho/\partial t=0$ boils down to a system of equations with $n$ unknowns, given by Eq.~{\eqref{eq:fullequation_Mrho10}}. We have solved these coupled equations in Appendix~{\ref{app:derivation_steady_n}} and discuss the example of a four-dimensional target system in Appendix~{\ref{app:steady_4}}.

Here, we clarify the benefit of the present approach which also deals with $n$-dimensional target systems. Essentially, this approach expands the idea of virtual qubits to compress the Hilbert space up to the size of the target system. For example, the Hilbert space of a $M$-level system coupled with $N$ two-qubit machines is $M\times 2^{2N}$. By applying the original idea of virtual qubit proposed by Ref.~{\cite{brunner2012virtual}} i.e. considering each two-qubit machine as one virtual qubit, the effective Hilbert space becomes $M\times 2^{N}$. However, the size of the effective space still grows exponentially at the number $N$ of virtual qubits. What we do further is to assume fast thermalisation rates such that all virtual qubits are thermalised at their virtual temperatures even more quickly than any other processes. This allows us to replace the virtual qubits with baths at their virtual temperatures, leading to compression of the Hilbert space to the size of the target system, i.e. $M$. The system size growth problem is solved by this. 

\subsection{Reset master equation (RME)} \label{sec:RME}

Here, we present the RME for the composite system (target and machine) and compare its steady state with the steady state~\eqref{eq:steady_3} of the effRME.
The RME describing the composite system reads
\begin{equation} \label{eq:master_reset}
    \frac{\partial \rho_{\mathrm{tot}}}{\partial t}
    =
    -i\left[H,\rho_{\mathrm{tot}}\right]
    +
    \sum_{i\in\mathcal{I}}
    Q_{i} \left(
    \tau_{i}\otimes \mathrm{Tr}_{i}[\rho_{\mathrm{tot}}] - \rho_{\mathrm{tot}}
    \right),
\end{equation}
with the respective thermalisation rates $Q_{i}$ for $\mathcal{I}\coloneq\{\mathrm{A}1,\mathrm{A}2,\mathrm{B}1,\mathrm{B}2,\mathrm{C}1,\mathrm{C}2\}$. The Hamiltonian is given by
\begin{align}\label{eq_H}
    H
    &=
    \sum_{k=0}^2\omega_k\ketbra{k}{k}
    +
    \sum_{i\in\mathcal{I}} \Omega_i\sigma^+_i\sigma_i^-
    +
    \big[
    g_\mathrm{A}\ketbra{0}{1}\sigma^+_{\mathrm{A}1}\sigma^-_{\mathrm{A}2}
    \nonumber\\
    &\quad
    +
    g_\mathrm{B}\ketbra{0}{2}\sigma^+_{\mathrm{B}1}\sigma^-_{\mathrm{B}2}
    +
    g_\mathrm{C}\ketbra{1}{2}\sigma^+_{\mathrm{C}1}\sigma^-_{\mathrm{C}2}
    +
    \mathrm{H.c.}
    \big]
\end{align}
with the coupling strengths $g_{\mathrm{A,B,C}}$ to each of the subsystems and the qubit frequencies $\Omega_{\mathrm{A}2}\coloneq \Omega_{\mathrm{A}1}-(\omega_1-\omega_0)$, $\Omega_{\mathrm{B}2}\coloneq \Omega_{\mathrm{B}1}-(\omega_2-\omega_0)$, $\Omega_{\mathrm{C}2}\coloneq \Omega_{\mathrm{C}1}-(\omega_2-\omega_1)$. While the solution of $\partial \rho_{\mathrm{tot}}/\partial t =0$ provides the steady state of the composite system, solving this equation analytically is difficult due to the size of the system, which is $3\times 2^2\times 2^2\times 2^2=192$. Even if numerical solutions of $\partial \rho_{\mathrm{tot}}/\partial t =0$ are obtained, it is hard to understand what the steady state shows physically and what kind of parameters characterise the steady state in contrast to the effRME.

In order to characterise the effective thermalisation rates $q_{\mathrm{A,B,C}}$ in the effRME, we compute the steady-state solution of the RME \eqref{eq:master_reset}.
Here, $\omega_0=0$ is taken, and the energy unit is set as half of the energy gap between the ground and first excited states of the qutrit, $(\omega_1-\omega_0)/2=\omega_1/2=1$.
Also, our focus is on a regime where the thermalisation rates $\{Q_{i}\}$ are much larger than any other energy scales such that the virtual temperatures are still valid.
By finding the population at each level in the RME solution corresponding to that in the effRME solution \eqref{eq:steady_3}, we have obtained the parameter dependency of the effective thermalisation rates $q_{\mathrm{A,B,C}}$. Assuming that all the two-qubit machines are subject to the same bath temperatures, i.e., $T_{i1}=T_{\mathrm{h}}$ and $T_{i2}=T_{\mathrm{c}}$ for $i\in\{\mathrm{A},\mathrm{B},\mathrm{C}\}$, we plot the ratio of the effective thermalisation rates $q_{\mathrm{A,B,C}}$ as a function of the coherent coupling strength $g_{\mathrm{B}}$ and the hot bath temperature $T_{\mathrm{h}}$ in Fig.~\ref{fig:ratioq_reset}. It is seen that the ratio $q_i/q_j$ is proportional to $g_i^2/g_j^2$ and that the $T_{\mathrm{h}}$-dependency of the ratio $q_i/q_j$ corresponds to the norm of virtual qubits, which, in analogy to Eq.~\eqref{eq:norm_vir}, reads
\begin{align}\label{eq_norm}
    n_{i}
    &=
    \tau_{i1}^{\mathrm{g}}\tau_{i2}^{\mathrm{e}}+\tau_{i1}^{\mathrm{e}}\tau_{i2}^{\mathrm{g}}
\end{align}
for $i\in\{\mathrm{A},\mathrm{B},\mathrm{C}\}$. 
These parameter dependencies are consistent with the effective thermalisation rate~\eqref{eq:qvir} in the case of a two-dimensional target system. 

Additionally, when $Q_{i1}=Q_{i2}=Q_i$, one can obtain the analytical form of the effective rates $q_i$ in the limit $Q_{i}\gg g_i$, particularly
\begin{align}  \label{eq:effectiveq_RME_text}
    q_i
    &
    =
    \frac{
    g_i^2
    \left(
    \tau_{i1}^{\mathrm{g}}\tau_{i2}^{\mathrm{e}}
    +
    \tau_{i1}^{\mathrm{e}}\tau_{i2}^{\mathrm{g}}
    \right)
    }{
    Q_{i}
    }.
\end{align}
See Appendix~{\ref{app:analyticalqj}} for the details.

\subsection{Gorini-Kossakowski-Lindblad-Sudarshan master equation (GKLSME)}\label{sec_GKLSME}

\begin{figure}
    \centering
    \includegraphics[width=0.98\columnwidth]{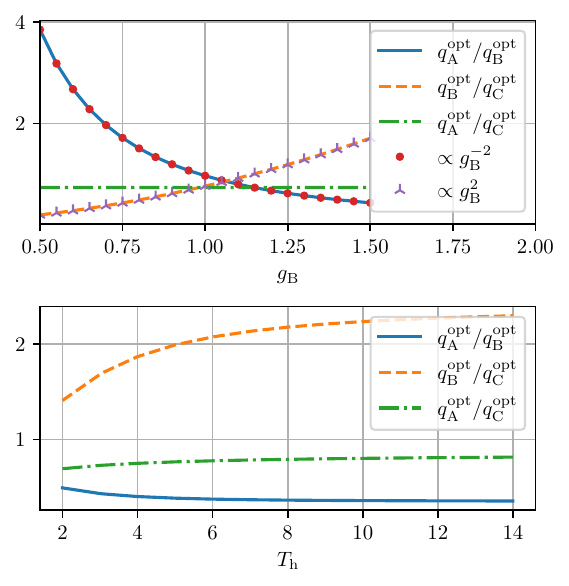}
    \caption{Optimal rates of the effRME~\eqref{eq:master_ABC} for which the state~\eqref{eq:steady_3} is the steady-state solution of the GKLSME~\eqref{eq_master_phys} as a function of the coherent coupling strength $g_\mathrm{B}$ (upper panel) and the hot-bath temperature $T_\mathrm{h}$ (lower panel). The quadratic relation~\eqref{eq_num_quadratic} can clearly be seen in the upper panel. 
    While $T_\mathrm{h}=3.1$ is used in the upper panel and $g_\mathrm{B}=1.5$ is used in the lower panel, both of the panels use the same values for the other parameters: 
    $\omega_0=0$, $\omega_1=2$, 
    $\omega_2=3$, $g_\mathrm{A}=1.2$, $g_\mathrm{C}=1.8$, $\Omega_{\mathrm{A}1}=2.5$, $\Omega_{\mathrm{B}1}=4.5$, $\Omega_{\mathrm{C}1}=1.3$, $T_\mathrm{c}=1.2$, $t=10$, $\Gamma_{i1}=70$, and $\Gamma_{i2}=50$ for $i\in\{\mathrm{A},\mathrm{B},\mathrm{C}\}$.
    In the same way as in Fig.~{\ref{fig:ratioq_reset}}, all the parameters are made dimensionless with the energy unit $(\omega_1-\omega_0)/2$.
    }
    \label{fig_numerics}
\end{figure}

We now continue to focus on the case of the target system being a qutrit. The idea of virtual qubits is then to replace the ``full'' RME~\eqref{eq:master_reset} that governs the dynamics of the joint system composed of the target qutrit and the six physical qubits (with the coupling rates $Q_i$) by an effRME that contains fewer ``virtual'' qubits (with the rates $q_i$). Namely, Eq.~\eqref{eq:master_reset} is replaced by Eq.~\eqref{eq:master_ABC}. The mapping of the rates, $\{Q_i\}\mapsto\{q_i\}$, is, in general, intricate. Notwithstanding, for the case of a target qubit we found that some features of the analytic relation~\eqref{eq:qvir}, such as the dependence on the coherent Hamiltonian coupling $g_i$ and the dependence on the norms, are numerically reproduced for a target qutrit in the foregoing section.

\par

The description of the physical system that underlies the effRME~\eqref{eq:master_ABC} is not unique and may depend on the concrete physical setup. In a sense, the RME~\eqref{eq:master_reset} constitutes, by itself, also an ad-hoc model, as it is not based on some continuous interaction with the environment, but a stochastic full swapping of constituent states with environment states. Nonetheless, owing to its CPTP (completely positive and trace-preserving) behaviour, the RME may be cast into a GKLSME, known from conventional thermalisation models~\cite{lindblad1976generators,gorini1976completely,breuerbook}. This mapping is, in general, a complicated function of the physical parameters, such as the bath temperatures, and has been explicitely derived for special cases~\cite{tavakoli2018heralded}. There, it is shown that in order to fulfil the mapping, the spontaneous emission rates $\Gamma_i$ in the GKLSME (see below) must be temperature-dependent, which is a feature that is usually not encountered in GKLSMEs~\cite{breuerbook}.

\par

On the other hand, we could have also chosen to formulate the original system in terms of a GKLSME with independent rates $\Gamma_i$ and then map it onto a ``full'' RME. Thereby, the rates $Q_i$ of the latter become themselves functions of the system parameters. As a consequence, the temperature dependence of the effective rates $q_i$ in the effRME for the target only is expected to depend on more than just the norms of the virtual qubits. The question of which description is more favourable depends on what parameters are easily tunable in a concrete experimental scenario. We will come back to this distinction in the next section and here assume the GKLSME to be the original equation and strive to understand the behaviour of $q_i$ in dependence of the physical parameters.

\par

The GKLSME for a target qutrit that interacts with six physical qubits reads
\begin{equation}\label{eq_master_phys}
  \frac{\partial\rho_\mathrm{tot}}{\partial t}
  =
  -i[H,\rho_\mathrm{tot}]+\sum_{i\in\mathcal{I}}\mathcal{L}_i\rho_\mathrm{tot},
\end{equation}
with the Hamiltonian~\eqref{eq_H} and the qubits $\mathcal{I}=\{\mathrm{A}1,\mathrm{A}2,\mathrm{B}1,\mathrm{B}2,\mathrm{C}1,\mathrm{C}2\}$. 
The Liouvillian
\begin{equation}
  \mathcal{L}_i\rho=\Gamma_i(\bar{n}(\Omega_i,T_i)+1)\mathcal{D}[\sigma^-_i]+\Gamma_i\bar{n}(\Omega_i,T_i)\mathcal{D}[\sigma^+_i]
\end{equation}
describes the dissipative interaction of the $i$th auxiliary qubit with its bath (see also Fig.~\ref{fig:qubit_vq_Ten}) at temperature $T_i=T_\mathrm{h}$ for $i\in\{\mathrm{A}1,\mathrm{B}1,\mathrm{C}1\}$ and $T_i=T_\mathrm{c}$ for $i\in\{\mathrm{A}2,\mathrm{B}2,\mathrm{C}2\}$, respectively; $\Gamma_i$ is that qubit's spontaneous emission rate. We have further defined the thermal population $\bar{n}(\omega,T)\coloneq [\exp(\omega/T)-1]^{-1}$ of the bosonic bath and the dissipator $D[A]\coloneq 2A\rho A^\dagger - A^\dagger A \rho - \rho A^\dagger A$.

\par

We now replace this equation with the simple effRME~\eqref{eq:master_ABC} for the qutrit only and pose the question: How are the parameters $\{q_\mathrm{A},q_\mathrm{B},q_\mathrm{C}\}$ of the effRME~\eqref{eq:master_ABC} related to the parameters of the GKLSME~\eqref{eq_master_phys}? To answer this question, we numerically integrate the GKLSME~\eqref{eq_master_phys} for given parameters with the analytic steady-state solution~\eqref{eq:steady_3}, which is parameterised by the triple $(q_\mathrm{A},q_\mathrm{B},q_\mathrm{C})$, as the initial state of the target qutrit (the qubits were initialised to their respective thermal states). We repeat this integration for different such triples to minimise the Frobenius norm $\|\rho(t)-\rho(0)\|$ between the reduced density operators of the qutrit at time $t$ and time $t=0$ for a sufficiently large fixed time $t>0$. In Fig.~{\ref{fig_numerics}} we chose $t=10$ to fulfill $t\gg 1/\Gamma_{i1,2}$. The Frobenius norm thus quantifies the deviation of the time-evolved state to the initial state~\eqref{eq:steady_3}. Thereby, we find the optimal parameter triple $(q_\mathrm{A}^\mathrm{opt},q_\mathrm{B}^\mathrm{opt},q_\mathrm{C}^\mathrm{opt})$ for which Eq.~\eqref{eq:steady_3} is the steady-state solution of Eq.~\eqref{eq_master_phys}. By repeating this procedure for, e.g., different $T_\mathrm{h}$, we can then numerically find the dependence of the rates $q_i$ on the physical parameters of the GKLSME (see Fig.~\ref{fig_numerics}).

\par

As seen from the upper panel in Fig.~\ref{fig_numerics}, the quadratic relation
\begin{equation}\label{eq_num_quadratic}
    \frac{q_i^\mathrm{opt}}{q_j^\mathrm{opt}}\propto \frac{g_i^2}{g_j^2}\text{ for }i,j\in \{\mathrm{A,B,C}\}
\end{equation}
of the effective rates in the effRME~\eqref{eq:master_ABC} to the Hamiltonian couplings in the GKLSME~\eqref{eq_master_phys}, first obtained in Eq.~\eqref{eq:qvir} for the qubit case, is reproduced, but with different proportionality factors than in Fig.~\ref{fig:ratioq_reset}. Whilst we only show the dependence on $g_\mathrm{B}$ in Fig.~\ref{fig_numerics}, we have performed additional simulations for varying $g_\mathrm{A}$ and $g_\mathrm{C}$, respectively, that are fully consistent with the quadratic behaviour in Eq.~\eqref{eq_num_quadratic}.
Furthermore, as expected, the temperature dependence does not agree with the norm of the virtual qubits (see the lower panel in Fig.~\ref{fig_numerics}), which will be discussed more in the next section.

In the limit $\Gamma_{i1},\Gamma_{i2}\gg g_i$, the analytical form of the effective rate $q_i$ can be obtained by using the Nakajima-Zwanzig projection operator technique~\cite{Erker2017,breuerbook},
\begin{align} \label{eq:effectiveq_GKLSME_text}
    q_{i}
    &=
    \frac{
    2 g_i^2
    \left(
    \tau_{i1}^{\mathrm{g}}\tau_{i2}^{\mathrm{e}}
    +
    \tau_{i1}^{\mathrm{e}}\tau_{i2}^{\mathrm{g}}
    \right)
    }{
    \Gamma_{i1}\left(\bar{n}(\Omega_{i1},T_{i1})+1\right)\mathcal{Z}_{i1}
    +
    \Gamma_{i2}\left(\bar{n}(\Omega_{i2},T_{i2})+1\right)\mathcal{Z}_{i2}
    }
    ,
\end{align}
where $\mathcal{Z}_{i1,2}=1+\mathrm{e}^{-\beta_{i1,2}\omega_{i}}$ are the partition functions. See Appendix~{\ref{app:analyticalqj}} for its derivation.

\par

We have implemented these simulations with the QuantumOptics.jl~\cite{kraemer2018quantumoptics} Julia framework and used Optim.jl~\cite{mogensen2018optim} for the numerical optimisation. The latter employed the Nelder--Mead method with threshold value $10^{-8}$. With the parameters of Fig.~{\ref{fig_numerics}}, the minimum of the Frobenius norm, $\|\rho(t)-\rho(0)\|_{q_\mathrm{opt}}$, then evaluated to $\sim 10^{-8}$.

\subsection{Discussion and identifying the ``dials''}

Our setup possesses a plethora of parameters, and the question on how they influence the steady-state solution of the target qutrit is not trivial. More importantly, to actually benefit from a simplification from the effRME it is important to understand how the physical parameters of different master equations impact the effRME. What we have seen above is that the behaviour of the optimal $q_i^\mathrm{opt}$ as a function of, e.g., the hot-bath temperature $T_\mathrm{h}$ differs, depending on the description: Whereas in the case of the reset model, the ratios of the $q_i^\mathrm{opt}$ depend on the corresponding ratio of the norms~\eqref{eq_norm} of the virtual qubits, this is not the case in the GKLS treatment.

\par

To understand this issue, it is important to note that in the RME for the seven-body system (target qutrit and six qubits) the parameters $Q_i$ were assumed to be \emph{independent} of the temperature. Therefore, the only temperature dependence in the $q_i^\mathrm{opt}$ stems from the norm. The GKLSME equivalent to the RME possesses temperature-dependent spontaneous emission rates. By contrast, in Sec.~\ref{sec_GKLSME} we have considered rates that do not depend on the temperatures. Therefore, the temperature dependence in Figs.~\ref{fig:ratioq_reset} and~\ref{fig_numerics} differ.

\par

It is important to note that although the RME can be written in GKLS form, the latter will not depict the behaviour that we are accustomed to from typical quantum-optical situations: Usually, the decay rates $\Gamma_i$ do not depend on the temperatures~\cite{breuerbook}, but the rates in the GKLS form of the reset equation will do, similar to Ref.~\cite{tavakoli2018heralded}. Therefore, features such as the temperature difference of the steady-state solution strongly depend on whether the $Q_i$ or the $\Gamma_i$ are assumed to be ``auxiliary'' parameters with no further dependence on the temperatures. If the $Q_i$ are deemed to be independent, then the $\Gamma_i$ will depend on the temperatures. Conversely, if the $\Gamma_i$ are chosen to be independent, then the $Q_i$ will depict a temperature dependence and would not correspond to the ratio of the norms anymore as shown in the lower panel in Fig.~\ref{fig_numerics}. We note that although the $\Gamma_i$ in the GKLS description typically depend on the frequencies~\cite{breuerbook}, we may still see them as independent parameters since the frequency dependence may be countered by, e.g., changing the dipole moment of the qubit. It is therefore sensible to assume the rates to be ``auxiliary'' parameters in either description (although the concrete parameter dependence of the ad-hoc $Q_i$ may be unknown). By contrast, the features that only depend on the Hamiltonian part of the master equation coincide in both descriptions, cf.\ the quadratic dependence on $g_\mathrm{B}$ in Figs.~\ref{fig:ratioq_reset} and~\ref{fig_numerics}, although the proportionality factors differ.

\par

It is therefore of paramount importance to distinguish between the two genuinely different models
\begin{itemize}
    \item RME with ``free'' $Q_i$
    \item GKLSME with ``free'' $\Gamma_i$
\end{itemize}
and the mapping of the RME to a GKLSME and vice versa, where the ``free'' character of the rates no longer holds. The ``dials'' therefore very much depend on the initial description of the machine, i.e., whether the $Q_i$ or the $\Gamma_i$ are supposed to be tunable by some auxiliary parameters. Both physical models, RME and GKLSME, therefore have their respective merit in different (experimental) setups.

\par

\section{Example: Improving a laser with population inversion}
\label{sec:laser_example}

\begin{figure}
    \centering
    \includegraphics[width=0.95\columnwidth]{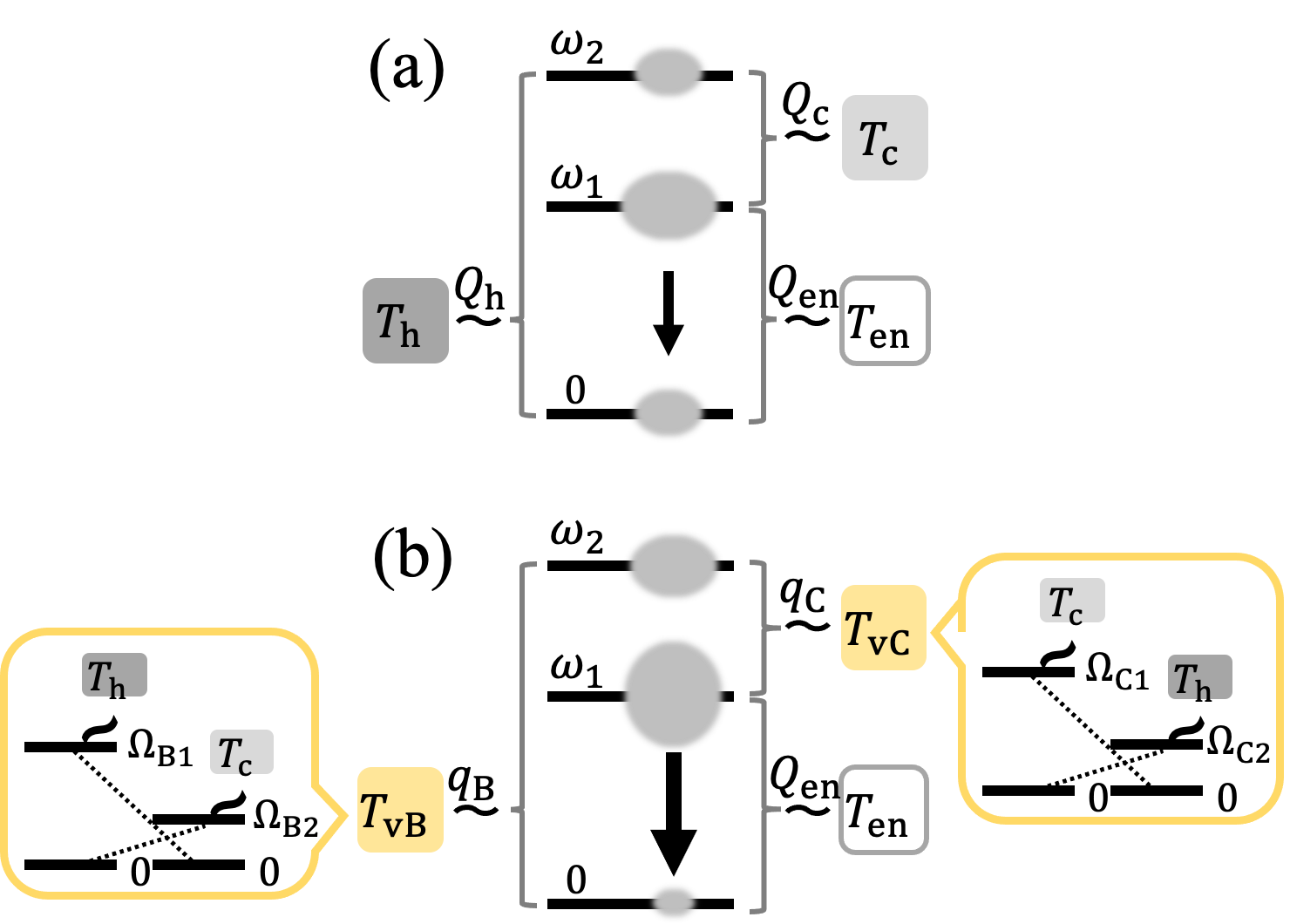}
    \caption{Sketches of (a) a typical laser mechanism with $T_{\mathrm{h}}$ the hot-bath temperature and $T_{\mathrm{c}}$ the cold-bath temperature, and (b) our proposed scheme improved by virtual temperatures, where $T_{\mathrm{vB}}$ [Eq.~\eqref{eq:TvB_laser}] is negative and $T_{\mathrm{vC}}$ [Eq.~\eqref{eq:TvC_laser}] is smaller than $T_{\mathrm{c}}$. The oval on each qutrit level represents this level's population. The lasing transition is further coupled to an environment at temperature $T_\mathrm{en}$.} 
    \label{fig:improve_laser}
\end{figure}


As an exemplary application of our method, we propose a scheme to enhance the output of a three-level laser by coupling it to a complex machine. A typical mechanism of a heat-pumped laser~\cite{scovil1959three,boukobza2007three,niedenzu2019concepts} is shown in Fig.~{\ref{fig:improve_laser}}(a). The laser is composed of a three-level system in contact with a hot bath at temperature $T_{\mathrm{h}}$, a cold bath at $T_{\mathrm{c}}$ and a signal field that is to be amplified. The lasing threshold is surpassed when the interactions with the two thermal baths generate a population inversion in the lasing transition $\ket{1}\rightarrow\ket{0}$ (black arrow in Fig.~{\ref{fig:improve_laser}}). 
The level structure considered in this section may be conducted with thermal atoms such as $^{87}$Rb (for example, see Ref.~{\cite{ghosh2018twolevel}}), and the hot and cold baths can be realised by thermal radiation filtered by narrow-band cavities~{\cite{ghosh2018twolevel}}.
Below, we address the question of whether this population inversion can be increased by indirectly coupling this three-level system to those temperatures via auxiliary two-qubit machines [Fig.~{\ref{fig:improve_laser}}(b)]. Notice that our interest is whether our approach improves an existing mechanism rather than how much inversions are generated based on our approach.



To address this question, we analyse the laser performance by computing the population inversion that is build up in a single lasing cycle, which allows computing the lasing threshold~\cite{scovil1959three}. We note that this approach has to be distinguished from the full quantum-mechanical treatment of the joint system composed of the three-level medium and the quantised light field~\cite{Boukobza2008,Youssef2009}. In this full treatment, the population inversion is ``cashed-in'' (utilised) to drive the light field into a Poissonian (phase-averaged coherent) state, which results in a significantly reduced remaining steady-state population inversion as compared to the cycle-based analysis~\cite{boukobza2006thermodynamic,niedenzu2019concepts}. Nevertheless, the lasing threshold and the lasing performance can already be deduced from our model. Thus, we consider only the laser's working medium below.



We replace the hot bath with a two-qubit machine whose virtual temperature $T_{\mathrm{vB}}$ is [cf.~Eq.~{\eqref{eq:Tv}}] 
\begin{align} \label{eq:TvB_laser}
    T_{\mathrm{vB}}
    &=
    \frac{\Omega_{\mathrm{B}1}-\Omega_{\mathrm{B}2}}{\Omega_{\mathrm{B}1}/T_{\mathrm{h}}-\Omega_{\mathrm{B}2}/T_{\mathrm{c}}}, 
\end{align}
where $\Omega_{\mathrm{B}2}=\Omega_{\mathrm{B}1}-(\omega_2-\omega_0)$. Note that for a fair comparison, the hot- and cold-bath temperatures used for the laser are also applied to this two-qubit machine. 
For $T_{\mathrm{h}}>(\Omega_{\mathrm{B1}}/\Omega_{\mathrm{B2}})T_{\mathrm{c}}$, the virtual temperature $T_{\mathrm{vB}}$ is negative and leads to population inversion between the levels $\ket{0}$ than $\ket{2}$ [see Fig.~\ref{fig:improve_laser}(b)]. 
Larger population in a higher energy state than a lower energy state is never seen with real thermal baths.
This population inversion between the levels $\ket{0}$ and $\ket{2}$ therefore increases the desired inversion on the lasing transition between $\ket{0}$ and $\ket{1}$ and hence increases the performance of the laser. 

\begin{figure}
    \centering
    \includegraphics[width=0.95\columnwidth]{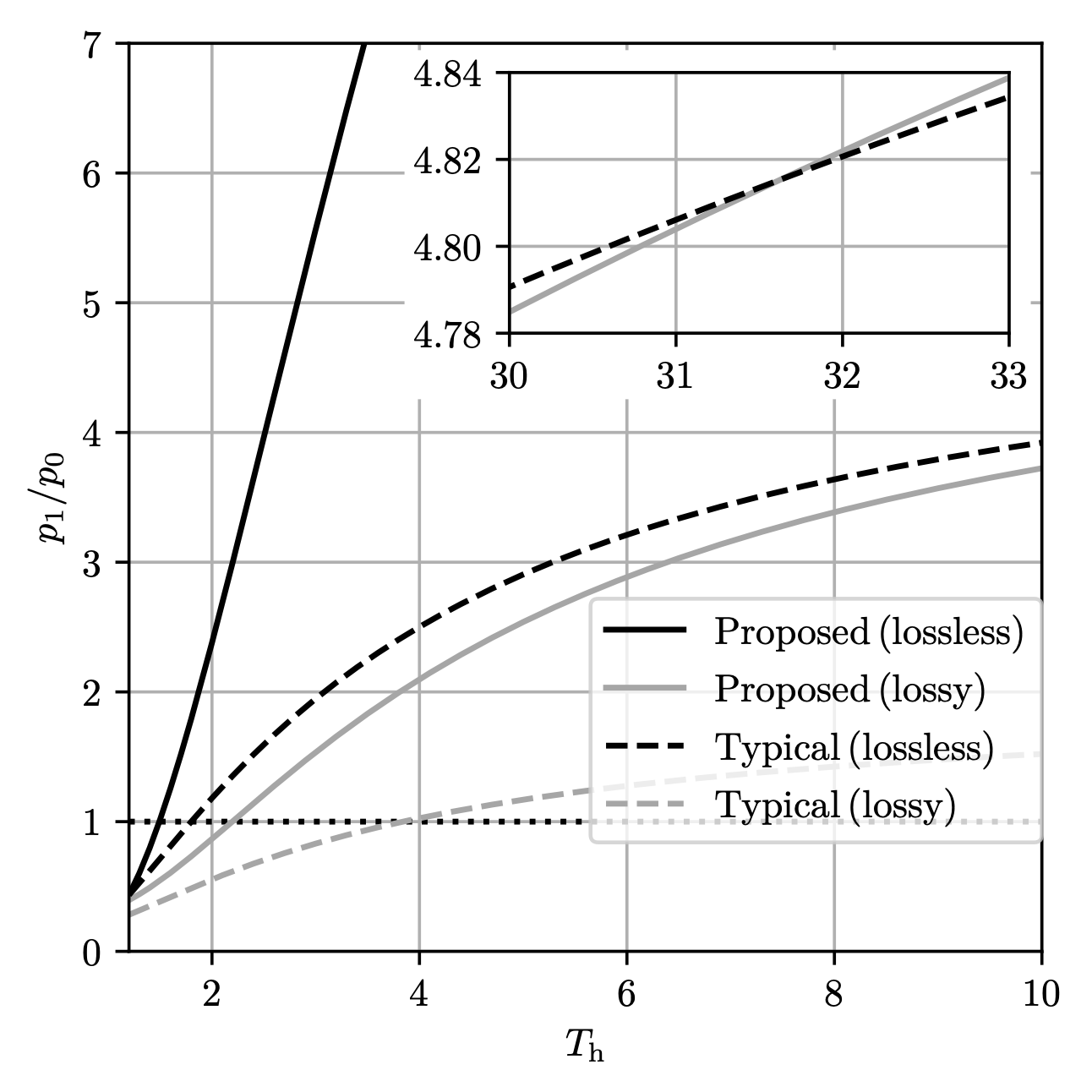}
    \caption{Population ratio $p_1/p_0$ of the levels $\ket{0}$ and $\ket{1}$ of the qutrit in the typical laser and our scheme when changing the hot temperature $T_{\mathrm{h}}$. The population ratio $p_1/p_0$ is displayed in cases of the typical laser (directly coupled to the heat baths) and our proposed scheme (indirectly coupled to the baths via two-qubit machines that give rise to virtual temperatures) with and without photon loss. The dotted black line represents the lasing threshold $p_1/p_0=1$. ``Lossless'' (``lossy'') means no (non-zero) photon loss. The inset zooms in on a regime that our proposed scheme with photon loss considered outperforms the lossless typical laser. The bath temperature and the thermalisation rate associated with loss of the laser output are assumed to be $T_{\mathrm{en}}=7.2$ and $Q_{\mathrm{en}}=0.1$. For the actual thermalisation rates, $Q_{\mathrm{h}}=2$ and $Q_{\mathrm{c}}=1.5$ are taken. The other parameters are the same as Figs.~{\ref{fig:ratioq_reset}} and~{\ref{fig_numerics}}, and all the parameters are made dimensionless with the energy unit $(\omega_1-\omega_0)/2$: $\omega_0=0$, $\omega_1=2$, $\omega_2=3$, $\Omega_{\mathrm{B}1}=4.5$, $\Omega_{\mathrm{C}1}=1.3$, $T_{\mathrm{c}}=1.2$.} 
    \label{fig:comparision_laser}
\end{figure}

The inversion on the lasing transition can be further improved by replacing the cold bath with a virtual qubit at the virtual temperature
\begin{align} \label{eq:TvC_laser}
    T_{\mathrm{vC}}
    &=
    \frac{\Omega_{\mathrm{C}1}-\Omega_{\mathrm{C}2}}{\Omega_{\mathrm{C}1}/T_{\mathrm{c}}-\Omega_{\mathrm{C}2}/T_{\mathrm{h}}}.
\end{align}
Since $T_{\mathrm{h}}>T_{\mathrm{c}}$ and $\Omega_{\mathrm{C}2}=\Omega_{\mathrm{C}1}-(\omega_2-\omega_1)$, $T_{\mathrm{vC}}$ is always lower than $T_{\mathrm{c}}$.



Hence, by tuning the qubit energies such that $T_{\mathrm{h}}>(\Omega_{\mathrm{B1}}/\Omega_{\mathrm{B2}})T_{\mathrm{c}}$, our proposed scheme works better than the typical laser in any parameter regime. For the ideal, i.e., lossless, case where the lasing transition is not subject to any additional environment, the optimal population inversion is realised if $T_{\mathrm{vB}}\to -0$ and $T_{\mathrm{vC}}\to +0$. For fixed bath temperatures, this could, e.g., be achieved by tuning the respective energy spacings $\Omega_{\mathrm{B}1}$ and $\Omega_{\mathrm{C}1}$ within the two-qubit machines.


For a more realistic situation with photon losses through an additional environment at temperature $T_\mathrm{en}$ that interacts with the lasing transition, the additional coupling $Q_{\mathrm{en}}$ competes with the rates to the virtual qubits. Namely, the steady state of the three-level system explicitly depends on those rates, and therefore the optimal virtual temperatures are no longer the same as the ideal lossless case. Accordingly, it is unclear whether the population ratio $p_1/p_0$ in the proposed scheme is still larger than in the typical scheme for fixed bath temperatures. Thus, we will compare $p_1/p_0$ in both of the schemes assuming photon losses.
Note that these additional losses on the lasing transition are not meant to describe the lasing process (which is a coherent process) but the action of any dissipative processes acting on the lasing transition. Those counteract the build-up of the population inversion and therefore reduce the lasing performance.

In analogy to Eq.~{\eqref{eq:steady_3}}, the steady state of the target qutrit reads
\begin{equation} \label{eq:steady_3_laser}
    \rho_{\mathrm{ss}}
    \propto
    \left(
    Q_{\mathrm{en}}\,
    q_{\mathrm{B}}\,
    \tau_{\mathrm{enB}}
    +
    q_{\mathrm{B}}\,
    q_{\mathrm{C}}\,
    \tau_{\mathrm{BC}}
    +
    q_{\mathrm{C}}\,
    Q_{\mathrm{en}}\,
    \tau_{\mathrm{Cen}}
    \right),
\end{equation}
where the states $\tau_{\mathrm{enB,BC,Cen}}$ are determined in analogy to Eqs.~\eqref{eq:tauABBCCA}. As seen in Figs.~\ref{fig:ratioq_reset} and~\ref{fig_numerics}, the rates $q_{\mathrm{B,C}}$ depend on the machine parameters. Therefore, these rates also change while tuning the bath temperatures $T_{\mathrm{h,c}}$ to control the virtual temperatures $T_{\mathrm{vB,vC}}$. By contrast, for the typical heat-pumped three-level laser, where the target is directly coupled to the thermal baths, the target relaxes to
\begin{equation} \label{eq:steady_3_typicallaser}
    \rho_{\mathrm{ss}}
    \propto
    \left(
    Q_{\mathrm{en}}\,
    Q_{\mathrm{h}}\,
    \tau_{\mathrm{enh}}
    +
    Q_{\mathrm{h}}\,
    Q_{\mathrm{c}}\,
    \tau_{\mathrm{hc}}
    +
    Q_{\mathrm{c}}\,
    Q_{\mathrm{en}}\,
    \tau_{\mathrm{cen}}
    \right),
\end{equation}
where the states $\tau_{\mathrm{enh,hc,cen}}$ are also obtained in analogy to Eqs.~\eqref{eq:tauABBCCA}.

To demonstrate that our proposed scheme can still outperform the typical three-level laser even in the non-ideal, lossy, case, we show the dependence of the population ratios $p_1/p_0$ obtained from Eq.~\eqref{eq:steady_3_laser} and Eq.~\eqref{eq:steady_3_typicallaser}, respectively, as a function of the hot-bath temperature in Fig.~\ref{fig:comparision_laser}. To this end, we choose fixed values for the couplings $Q_\mathrm{h}$, $Q_\mathrm{c}$ and $Q_\mathrm{en}$ and set the effective rates to be
\begin{subequations}\label{eq_effective_rates_plot_laser}
\begin{align}
    q_\mathrm{B}&=Q_\mathrm{h}n_\mathrm{B},\\
    q_\mathrm{C}&=Q_\mathrm{c}n_\mathrm{C},
\end{align}
\end{subequations}
with the norms~\eqref{eq_norm} to allow for a fair comparison between the two laser setups in Fig.~\ref{fig:improve_laser}. Namely, we assume the internal details of the two-qubit machines [yellow boxes in Fig.~\ref{fig:improve_laser}(b)] to be tuned in such a way that the effective rates are Eqs.~\eqref{eq_effective_rates_plot_laser}. 
This is just a choice that works as shown below. If one wants to control the system to obtain a desirable state, what one needs to know is the parameter dependence. Since one knows the interaction- and temperature- dependence of the effective rate, one finds a set of desirable interaction strengths and temperatures.

It is seen in Fig.~\ref{fig:comparision_laser} that using the more complex setup with virtual qubits the population inversion of the lasing transition can be strongly increased.
For the chosen parameters, our scheme allowing for photon loss outperforms even the typical laser outcome in the ideal case for $T_\mathrm{h}\gtrsim 32$ (see the inset in Fig.~\ref{fig:comparision_laser}). Note that we only tune $T_{\mathrm{h}}$ and leave the other parameters such as the qubit frequencies invariant. If one tunes the other parameters as well, the lasing transition can be improved efficiently. 

\par

\section{Conclusions}

Designing complex thermal machines at the quantum scale is hard, as they quickly become intractable. We have instead decided to model only the steady state of an arbitrary-dimensional target system in contact with complex machinery coupled to different heat baths. This can be done by means of competing virtual qubits, coupled to the different transitions of the quantum target. Using reset-type master equations enables one to have an analytical solution for all dimensions. We have studied and showcased the behaviour in three dimension, comparing it to full solutions of an optical master equation (GKLSME) and showed that they share central features and behaviours, whereas the exact target state can at times be different. 
Furthermore, we have displayed that the parameters in the effRME can be actually determined analytically when starting a GKLSME or a RME. The analytical form of $q_i$ enables one to predict the parameter dependence of the steady state.
We believe that the models prove usefulness for designing machines to optimise certain key properties of the target system, such as inverting the population of the lasing transition in a three-level laser, or generally to optimally create purity in a subspace of the multi-level system. 

Based on our results, there are two possible applications: (i) given a qudit coupled to some two-qubit machines and a master equation describing this system, one can derive the steady state of the qudit; (ii) given a desired steady  state of a qudit, one can design the parameters in the machines to produce it. The first option~(i) is possible by using the analytical form of the steady state~{\eqref{eq:steady_n}} and the effective rates~{\eqref{eq:effectiveq_RME_text}} or {\eqref{eq:effectiveq_GKLSME_text}} (or deriving a suitable effective rate with the technique presented in Appendix~{\ref{app:analyticalqj}}).
The second option~(ii) is also achievable. For any $n$-level system, the number of variables for the population is $n-1$ due to the normalisation. By coupling at least $n-1$ two-qubit machines to all the level of the $n$-level system, one can control all the populations and find a desirable effRME. Note that it is not necessary to couple machines to all the $n(n-1)/2$ transition of the $n$-level system. For example, coupling $n-1$ machines to neighbouring transitions is sufficient to determine the population in the $n$-level system, and each transition will thermalise at the virtual temperature in the asymptotic limit. This is completely independent of interaction strengths, and applying our results is not necessary. However, our results become useful when multiple transitions couple to the same level, creating competitions between the different rates. This is a pertinent scenario as shown in Sec.~{\ref{sec:laser_example}}.

One of the potential applications is modelling highly complex open systems that are not amenable to a full GKLSME solution, either due to size or unknown Hamiltonian parameters. While in such a situation, it is always possible to fit an effRME to experimental data, the real challenge is an understanding of how physical control over temperatures and couplings will impact the parameters of the effRME. While in practice one can expect that this could be fitted through many experimental runs in different parameter regimes, to endow the effRME with predictive power it would be great to get more qualitative insight into parameter correspondences. We have seen that for qutrits the simplified model parameters can easily be matched to physical parameters, either quantitatively or qualitatively. This correspondence can be seen in higher dimensional systems by using the same technique presented in Appendix~{\ref{app:analyticalqj}}.

\section*{Acknowledgments}
A.\,U.\ appreciates the hospitality and support from IQOQI-Vienna during her visit and acknowledges financial support from OIST Graduate University, Research Fellowship of JSPS for Young Scientists, and JSPS KAKENHI Grant Number 20J10006. W.\,N.\ acknowledges support from an ESQ fellowship of the Austrian Academy of Sciences (\"OAW). M.~H.~ also acknowledges funds from the FQXi (FQXi-IAF19-03-S2)
 within the project ``\emph{Fueling quantum field machines with information}'' and from the Austrian Science Fund (FWF) through the START project Y879-N27. All authors acknowledge productive discussions with the QUIT physics group and useful comments on the manuscript from Thomas Busch. 

\bibliographystyle{apsrev4-1-title}
\bibliography{virtual}

\begin{thebibliography}{41}%
\makeatletter
\providecommand \@ifxundefined [1]{%
 \@ifx{#1\undefined}
}%
\providecommand \@ifnum [1]{%
 \ifnum #1\expandafter \@firstoftwo
 \else \expandafter \@secondoftwo
 \fi
}%
\providecommand \@ifx [1]{%
 \ifx #1\expandafter \@firstoftwo
 \else \expandafter \@secondoftwo
 \fi
}%
\providecommand \natexlab [1]{#1}%
\providecommand \enquote  [1]{#1}%
\providecommand \bibnamefont  [1]{#1}%
\providecommand \bibfnamefont [1]{#1}%
\providecommand \citenamefont [1]{#1}%
\providecommand \href@noop [0]{\@secondoftwo}%
\providecommand \href [0]{\begingroup \@sanitize@url \@href}%
\providecommand \@href[1]{\@@startlink{#1}\@@href}%
\providecommand \@@href[1]{\endgroup#1\@@endlink}%
\providecommand \@sanitize@url [0]{\catcode `\\12\catcode `\$12\catcode
  `\&12\catcode `\#12\catcode `\^12\catcode `\_12\catcode `\%12\relax}%
\providecommand \@@startlink[1]{}%
\providecommand \@@endlink[0]{}%
\providecommand \url  [0]{\begingroup\@sanitize@url \@url }%
\providecommand \@url [1]{\endgroup\@href {#1}{\urlprefix }}%
\providecommand \urlprefix  [0]{URL }%
\providecommand \Eprint [0]{\href }%
\providecommand \doibase [0]{http://dx.doi.org/}%
\providecommand \selectlanguage [0]{\@gobble}%
\providecommand \bibinfo  [0]{\@secondoftwo}%
\providecommand \bibfield  [0]{\@secondoftwo}%
\providecommand \translation [1]{[#1]}%
\providecommand \BibitemOpen [0]{}%
\providecommand \bibitemStop [0]{}%
\providecommand \bibitemNoStop [0]{.\EOS\space}%
\providecommand \EOS [0]{\spacefactor3000\relax}%
\providecommand \BibitemShut  [1]{\csname bibitem#1\endcsname}%
\let\auto@bib@innerbib\@empty
\bibitem [{\citenamefont {Alicki}(1979)}]{alicki1979quantum}%
  \BibitemOpen
  \bibfield  {author} {\bibinfo {author} {\bibfnamefont {R.}~\bibnamefont
  {Alicki}},\ }\enquote {\bibinfo {title} {The quantum open system as a model
  of the heat engine},}\ \href {\doibase 10.1088/0305-4470/12/5/007} {\bibfield
   {journal} {\bibinfo  {journal} {J. Phys. A}\ }\textbf {\bibinfo {volume}
  {12}},\ \bibinfo {pages} {L103} (\bibinfo {year} {1979})}\BibitemShut
  {NoStop}%
\bibitem [{\citenamefont {Kosloff}(1984)}]{kosloff1984quantum}%
  \BibitemOpen
  \bibfield  {author} {\bibinfo {author} {\bibfnamefont {R.}~\bibnamefont
  {Kosloff}},\ }\enquote {\bibinfo {title} {A quantum mechanical open system as
  a model of a heat engine},}\ \href {\doibase 10.1063/1.446862} {\bibfield
  {journal} {\bibinfo  {journal} {J. Chem. Phys.}\ }\textbf {\bibinfo {volume}
  {80}},\ \bibinfo {pages} {1625} (\bibinfo {year} {1984})}\BibitemShut
  {NoStop}%
\bibitem [{\citenamefont {Kosloff}(2013)}]{kosloff2013quantum}%
  \BibitemOpen
  \bibfield  {author} {\bibinfo {author} {\bibfnamefont {R.}~\bibnamefont
  {Kosloff}},\ }\enquote {\bibinfo {title} {Quantum Thermodynamics: A Dynamical
  Viewpoint},}\ \href {\doibase 10.3390/e15062100} {\bibfield  {journal}
  {\bibinfo  {journal} {Entropy}\ }\textbf {\bibinfo {volume} {15}},\ \bibinfo
  {pages} {2100} (\bibinfo {year} {2013})}\BibitemShut {NoStop}%
\bibitem [{\citenamefont {Gelbwaser-Klimovsky}\ \emph
  {et~al.}(2015)\citenamefont {Gelbwaser-Klimovsky}, \citenamefont {Niedenzu},\
  and\ \citenamefont {Kurizki}}]{gelbwaser2015thermodynamics}%
  \BibitemOpen
  \bibfield  {author} {\bibinfo {author} {\bibfnamefont {D.}~\bibnamefont
  {Gelbwaser-Klimovsky}}, \bibinfo {author} {\bibfnamefont {W.}~\bibnamefont
  {Niedenzu}}, \ and\ \bibinfo {author} {\bibfnamefont {G.}~\bibnamefont
  {Kurizki}},\ }\enquote {\bibinfo {title} {Thermodynamics of Quantum Systems
  Under Dynamical Control},}\ \href {\doibase 10.1016/bs.aamop.2015.07.002}
  {\bibfield  {journal} {\bibinfo  {journal} {Adv. At. Mol. Opt. Phys.}\
  }\textbf {\bibinfo {volume} {64}},\ \bibinfo {pages} {329} (\bibinfo {year}
  {2015})}\BibitemShut {NoStop}%
\bibitem [{\citenamefont {Goold}\ \emph {et~al.}(2016)\citenamefont {Goold},
  \citenamefont {Huber}, \citenamefont {Riera}, \citenamefont {del Rio},\ and\
  \citenamefont {Skrzypczyk}}]{goold2016role}%
  \BibitemOpen
  \bibfield  {author} {\bibinfo {author} {\bibfnamefont {J.}~\bibnamefont
  {Goold}}, \bibinfo {author} {\bibfnamefont {M.}~\bibnamefont {Huber}},
  \bibinfo {author} {\bibfnamefont {A.}~\bibnamefont {Riera}}, \bibinfo
  {author} {\bibfnamefont {L.}~\bibnamefont {del Rio}}, \ and\ \bibinfo
  {author} {\bibfnamefont {P.}~\bibnamefont {Skrzypczyk}},\ }\enquote {\bibinfo
  {title} {The role of quantum information in thermodynamics---a topical
  review},}\ \href {\doibase 10.1088/1751-8113/49/14/143001} {\bibfield
  {journal} {\bibinfo  {journal} {J. Phys. A}\ }\textbf {\bibinfo {volume}
  {49}},\ \bibinfo {pages} {143001} (\bibinfo {year} {2016})}\BibitemShut
  {NoStop}%
\bibitem [{\citenamefont {Vinjanampathy}\ and\ \citenamefont
  {Anders}(2016)}]{vinjanampathy2016quantum}%
  \BibitemOpen
  \bibfield  {author} {\bibinfo {author} {\bibfnamefont {S.}~\bibnamefont
  {Vinjanampathy}}\ and\ \bibinfo {author} {\bibfnamefont {J.}~\bibnamefont
  {Anders}},\ }\enquote {\bibinfo {title} {Quantum thermodynamics},}\ \href
  {\doibase 10.1080/00107514.2016.1201896} {\bibfield  {journal} {\bibinfo
  {journal} {Contemp. Phys.}\ }\textbf {\bibinfo {volume} {57}},\ \bibinfo
  {pages} {545} (\bibinfo {year} {2016})}\BibitemShut {NoStop}%
\bibitem [{\citenamefont {Ghosh}\ \emph {et~al.}(2019)\citenamefont {Ghosh},
  \citenamefont {Niedenzu}, \citenamefont {Mukherjee},\ and\ \citenamefont
  {Kurizki}}]{ghosh2019thermodynamic}%
  \BibitemOpen
  \bibfield  {author} {\bibinfo {author} {\bibfnamefont {A.}~\bibnamefont
  {Ghosh}}, \bibinfo {author} {\bibfnamefont {W.}~\bibnamefont {Niedenzu}},
  \bibinfo {author} {\bibfnamefont {V.}~\bibnamefont {Mukherjee}}, \ and\
  \bibinfo {author} {\bibfnamefont {G.}~\bibnamefont {Kurizki}},\ }\enquote
  {\bibinfo {title} {Thermodynamic Principles and Implementations of Quantum
  Machines},}\ in\ \href {\doibase 10.1007/978-3-319-99046-0_2} {\emph
  {\bibinfo {booktitle} {Thermodynamics in the Quantum Regime}}},\ \bibinfo
  {editor} {edited by\ \bibinfo {editor} {\bibfnamefont {F.}~\bibnamefont
  {Binder}}, \bibinfo {editor} {\bibfnamefont {L.~A.}\ \bibnamefont {Correa}},
  \bibinfo {editor} {\bibfnamefont {C.}~\bibnamefont {Gogolin}}, \bibinfo
  {editor} {\bibfnamefont {J.}~\bibnamefont {Anders}}, \ and\ \bibinfo {editor}
  {\bibfnamefont {G.}~\bibnamefont {Adesso}}}\ (\bibinfo  {publisher}
  {Springer},\ \bibinfo {address} {Cham},\ \bibinfo {year} {2019})\ pp.\
  \bibinfo {pages} {37--66}\BibitemShut {NoStop}%
\bibitem [{\citenamefont {Bhattacharjee}\ and\ \citenamefont
  {Dutta}(2020)}]{bhattacharjee2020quantumthermal}%
  \BibitemOpen
  \bibfield  {author} {\bibinfo {author} {\bibfnamefont {S.}~\bibnamefont
  {Bhattacharjee}}\ and\ \bibinfo {author} {\bibfnamefont {A.}~\bibnamefont
  {Dutta}},\ }\enquote {\bibinfo {title} {Quantum thermal machines and
  batteries},}\ \href {https://arxiv.org/abs/2008.07889} {\bibfield  {journal}
  {\bibinfo  {journal} {arXiv preprint arXiv:2008.07889}\ } (\bibinfo {year}
  {2020})}\BibitemShut {NoStop}%
\bibitem [{\citenamefont {Koski}\ \emph {et~al.}(2014)\citenamefont {Koski},
  \citenamefont {Maisi}, \citenamefont {Pekola},\ and\ \citenamefont
  {Averin}}]{koski2014experimental}%
  \BibitemOpen
  \bibfield  {author} {\bibinfo {author} {\bibfnamefont {J.~V.}\ \bibnamefont
  {Koski}}, \bibinfo {author} {\bibfnamefont {V.~F.}\ \bibnamefont {Maisi}},
  \bibinfo {author} {\bibfnamefont {J.~P.}\ \bibnamefont {Pekola}}, \ and\
  \bibinfo {author} {\bibfnamefont {D.~V.}\ \bibnamefont {Averin}},\ }\enquote
  {\bibinfo {title} {Experimental realization of a Szilard engine with a single
  electron},}\ \href {\doibase 10.1073/pnas.1406966111} {\bibfield  {journal}
  {\bibinfo  {journal} {Proc. Natl. Acad. Sci. USA}\ }\textbf {\bibinfo
  {volume} {111}},\ \bibinfo {pages} {13786} (\bibinfo {year}
  {2014})}\BibitemShut {NoStop}%
\bibitem [{\citenamefont {Ro{\ss}nagel}\ \emph {et~al.}(2016)\citenamefont
  {Ro{\ss}nagel}, \citenamefont {Dawkins}, \citenamefont {Tolazzi},
  \citenamefont {Abah}, \citenamefont {Lutz}, \citenamefont {Schmidt-Kaler},\
  and\ \citenamefont {Singer}}]{rossnagel2016single}%
  \BibitemOpen
  \bibfield  {author} {\bibinfo {author} {\bibfnamefont {J.}~\bibnamefont
  {Ro{\ss}nagel}}, \bibinfo {author} {\bibfnamefont {S.~T.}\ \bibnamefont
  {Dawkins}}, \bibinfo {author} {\bibfnamefont {K.~N.}\ \bibnamefont
  {Tolazzi}}, \bibinfo {author} {\bibfnamefont {O.}~\bibnamefont {Abah}},
  \bibinfo {author} {\bibfnamefont {E.}~\bibnamefont {Lutz}}, \bibinfo {author}
  {\bibfnamefont {F.}~\bibnamefont {Schmidt-Kaler}}, \ and\ \bibinfo {author}
  {\bibfnamefont {K.}~\bibnamefont {Singer}},\ }\enquote {\bibinfo {title} {A
  single-atom heat engine},}\ \href {\doibase 10.1126/science.aad6320}
  {\bibfield  {journal} {\bibinfo  {journal} {Science}\ }\textbf {\bibinfo
  {volume} {352}},\ \bibinfo {pages} {325} (\bibinfo {year}
  {2016})}\BibitemShut {NoStop}%
\bibitem [{\citenamefont {Klaers}\ \emph {et~al.}(2017)\citenamefont {Klaers},
  \citenamefont {Faelt}, \citenamefont {Imamoglu},\ and\ \citenamefont
  {Togan}}]{klaers2017squeezed}%
  \BibitemOpen
  \bibfield  {author} {\bibinfo {author} {\bibfnamefont {J.}~\bibnamefont
  {Klaers}}, \bibinfo {author} {\bibfnamefont {S.}~\bibnamefont {Faelt}},
  \bibinfo {author} {\bibfnamefont {A.}~\bibnamefont {Imamoglu}}, \ and\
  \bibinfo {author} {\bibfnamefont {E.}~\bibnamefont {Togan}},\ }\enquote
  {\bibinfo {title} {Squeezed Thermal Reservoirs as a Resource for a
  Nanomechanical Engine beyond the Carnot Limit},}\ \href {\doibase
  10.1103/PhysRevX.7.031044} {\bibfield  {journal} {\bibinfo  {journal} {Phys.
  Rev. X}\ }\textbf {\bibinfo {volume} {7}},\ \bibinfo {pages} {031044}
  (\bibinfo {year} {2017})}\BibitemShut {NoStop}%
\bibitem [{\citenamefont {Peterson}\ \emph {et~al.}(2019)\citenamefont
  {Peterson}, \citenamefont {Batalh\~ao}, \citenamefont {Herrera},
  \citenamefont {Souza}, \citenamefont {Sarthour}, \citenamefont {Oliveira},\
  and\ \citenamefont {Serra}}]{peterson2019experimental}%
  \BibitemOpen
  \bibfield  {author} {\bibinfo {author} {\bibfnamefont {J.~P.~S.}\
  \bibnamefont {Peterson}}, \bibinfo {author} {\bibfnamefont {T.~B.}\
  \bibnamefont {Batalh\~ao}}, \bibinfo {author} {\bibfnamefont
  {M.}~\bibnamefont {Herrera}}, \bibinfo {author} {\bibfnamefont {A.~M.}\
  \bibnamefont {Souza}}, \bibinfo {author} {\bibfnamefont {R.~S.}\ \bibnamefont
  {Sarthour}}, \bibinfo {author} {\bibfnamefont {I.~S.}\ \bibnamefont
  {Oliveira}}, \ and\ \bibinfo {author} {\bibfnamefont {R.~M.}\ \bibnamefont
  {Serra}},\ }\enquote {\bibinfo {title} {Experimental Characterization of a
  Spin Quantum Heat Engine},}\ \href {\doibase 10.1103/PhysRevLett.123.240601}
  {\bibfield  {journal} {\bibinfo  {journal} {Phys. Rev. Lett.}\ }\textbf
  {\bibinfo {volume} {123}},\ \bibinfo {pages} {240601} (\bibinfo {year}
  {2019})}\BibitemShut {NoStop}%
\bibitem [{\citenamefont {von Lindenfels}\ \emph {et~al.}(2019)\citenamefont
  {von Lindenfels}, \citenamefont {Gr\"ab}, \citenamefont {Schmiegelow},
  \citenamefont {Kaushal}, \citenamefont {Schulz}, \citenamefont {Mitchison},
  \citenamefont {Goold}, \citenamefont {Schmidt-Kaler},\ and\ \citenamefont
  {Poschinger}}]{vonlindenfels2019spin}%
  \BibitemOpen
  \bibfield  {author} {\bibinfo {author} {\bibfnamefont {D.}~\bibnamefont {von
  Lindenfels}}, \bibinfo {author} {\bibfnamefont {O.}~\bibnamefont {Gr\"ab}},
  \bibinfo {author} {\bibfnamefont {C.~T.}\ \bibnamefont {Schmiegelow}},
  \bibinfo {author} {\bibfnamefont {V.}~\bibnamefont {Kaushal}}, \bibinfo
  {author} {\bibfnamefont {J.}~\bibnamefont {Schulz}}, \bibinfo {author}
  {\bibfnamefont {M.~T.}\ \bibnamefont {Mitchison}}, \bibinfo {author}
  {\bibfnamefont {J.}~\bibnamefont {Goold}}, \bibinfo {author} {\bibfnamefont
  {F.}~\bibnamefont {Schmidt-Kaler}}, \ and\ \bibinfo {author} {\bibfnamefont
  {U.~G.}\ \bibnamefont {Poschinger}},\ }\enquote {\bibinfo {title} {Spin Heat
  Engine Coupled to a Harmonic-Oscillator Flywheel},}\ \href {\doibase
  10.1103/PhysRevLett.123.080602} {\bibfield  {journal} {\bibinfo  {journal}
  {Phys. Rev. Lett.}\ }\textbf {\bibinfo {volume} {123}},\ \bibinfo {pages}
  {080602} (\bibinfo {year} {2019})}\BibitemShut {NoStop}%
\bibitem [{\citenamefont {Klatzow}\ \emph {et~al.}(2019)\citenamefont
  {Klatzow}, \citenamefont {Becker}, \citenamefont {Ledingham}, \citenamefont
  {Weinzetl}, \citenamefont {Kaczmarek}, \citenamefont {Saunders},
  \citenamefont {Nunn}, \citenamefont {Walmsley}, \citenamefont {Uzdin},\ and\
  \citenamefont {Poem}}]{klatzow2019experimental}%
  \BibitemOpen
  \bibfield  {author} {\bibinfo {author} {\bibfnamefont {J.}~\bibnamefont
  {Klatzow}}, \bibinfo {author} {\bibfnamefont {J.~N.}\ \bibnamefont {Becker}},
  \bibinfo {author} {\bibfnamefont {P.~M.}\ \bibnamefont {Ledingham}}, \bibinfo
  {author} {\bibfnamefont {C.}~\bibnamefont {Weinzetl}}, \bibinfo {author}
  {\bibfnamefont {K.~T.}\ \bibnamefont {Kaczmarek}}, \bibinfo {author}
  {\bibfnamefont {D.~J.}\ \bibnamefont {Saunders}}, \bibinfo {author}
  {\bibfnamefont {J.}~\bibnamefont {Nunn}}, \bibinfo {author} {\bibfnamefont
  {I.~A.}\ \bibnamefont {Walmsley}}, \bibinfo {author} {\bibfnamefont
  {R.}~\bibnamefont {Uzdin}}, \ and\ \bibinfo {author} {\bibfnamefont
  {E.}~\bibnamefont {Poem}},\ }\enquote {\bibinfo {title} {Experimental
  Demonstration of Quantum Effects in the Operation of Microscopic Heat
  Engines},}\ \href {\doibase 10.1103/PhysRevLett.122.110601} {\bibfield
  {journal} {\bibinfo  {journal} {Phys. Rev. Lett.}\ }\textbf {\bibinfo
  {volume} {122}},\ \bibinfo {pages} {110601} (\bibinfo {year}
  {2019})}\BibitemShut {NoStop}%
\bibitem [{\citenamefont {Brunner}\ \emph {et~al.}(2012)\citenamefont
  {Brunner}, \citenamefont {Linden}, \citenamefont {Popescu},\ and\
  \citenamefont {Skrzypczyk}}]{brunner2012virtual}%
  \BibitemOpen
  \bibfield  {author} {\bibinfo {author} {\bibfnamefont {N.}~\bibnamefont
  {Brunner}}, \bibinfo {author} {\bibfnamefont {N.}~\bibnamefont {Linden}},
  \bibinfo {author} {\bibfnamefont {S.}~\bibnamefont {Popescu}}, \ and\
  \bibinfo {author} {\bibfnamefont {P.}~\bibnamefont {Skrzypczyk}},\ }\enquote
  {\bibinfo {title} {Virtual qubits, virtual temperatures, and the foundations
  of thermodynamics},}\ \href {\doibase 10.1103/PhysRevE.85.051117} {\bibfield
  {journal} {\bibinfo  {journal} {Phys. Rev. E}\ }\textbf {\bibinfo {volume}
  {85}},\ \bibinfo {pages} {051117} (\bibinfo {year} {2012})}\BibitemShut
  {NoStop}%
\bibitem [{\citenamefont {Linden}\ \emph {et~al.}(2010)\citenamefont {Linden},
  \citenamefont {Popescu},\ and\ \citenamefont {Skrzypczyk}}]{Linden2010}%
  \BibitemOpen
  \bibfield  {author} {\bibinfo {author} {\bibfnamefont {N.}~\bibnamefont
  {Linden}}, \bibinfo {author} {\bibfnamefont {S.}~\bibnamefont {Popescu}}, \
  and\ \bibinfo {author} {\bibfnamefont {P.}~\bibnamefont {Skrzypczyk}},\
  }\enquote {\bibinfo {title} {How Small Can Thermal Machines Be? The Smallest
  Possible Refrigerator},}\ \href {\doibase 10.1103/PhysRevLett.105.130401}
  {\bibfield  {journal} {\bibinfo  {journal} {Phys. Rev. Lett.}\ }\textbf
  {\bibinfo {volume} {105}},\ \bibinfo {pages} {130401} (\bibinfo {year}
  {2010})}\BibitemShut {NoStop}%
\bibitem [{\citenamefont {Skrzypczyk}\ \emph {et~al.}(2011)\citenamefont
  {Skrzypczyk}, \citenamefont {Brunner}, \citenamefont {Linden},\ and\
  \citenamefont {Popescu}}]{skrzypczyk2011smallest}%
  \BibitemOpen
  \bibfield  {author} {\bibinfo {author} {\bibfnamefont {P.}~\bibnamefont
  {Skrzypczyk}}, \bibinfo {author} {\bibfnamefont {N.}~\bibnamefont {Brunner}},
  \bibinfo {author} {\bibfnamefont {N.}~\bibnamefont {Linden}}, \ and\ \bibinfo
  {author} {\bibfnamefont {S.}~\bibnamefont {Popescu}},\ }\enquote {\bibinfo
  {title} {The smallest refrigerators can reach maximal efficiency},}\ \href
  {\doibase 10.1088/1751-8113/44/49/492002} {\bibfield  {journal} {\bibinfo
  {journal} {J. Phys. A: Math. Theor.}\ }\textbf {\bibinfo {volume} {44}},\
  \bibinfo {pages} {492002} (\bibinfo {year} {2011})}\BibitemShut {NoStop}%
\bibitem [{\citenamefont {Silva}\ \emph {et~al.}(2016)\citenamefont {Silva},
  \citenamefont {Manzano}, \citenamefont {Skrzypczyk},\ and\ \citenamefont
  {Brunner}}]{Silva2016}%
  \BibitemOpen
  \bibfield  {author} {\bibinfo {author} {\bibfnamefont {R.}~\bibnamefont
  {Silva}}, \bibinfo {author} {\bibfnamefont {G.}~\bibnamefont {Manzano}},
  \bibinfo {author} {\bibfnamefont {P.}~\bibnamefont {Skrzypczyk}}, \ and\
  \bibinfo {author} {\bibfnamefont {N.}~\bibnamefont {Brunner}},\ }\enquote
  {\bibinfo {title} {Performance of autonomous quantum thermal machines:
  Hilbert space dimension as a thermodynamical resource},}\ \href {\doibase
  10.1103/PhysRevE.94.032120} {\bibfield  {journal} {\bibinfo  {journal} {Phys.
  Rev. E}\ }\textbf {\bibinfo {volume} {94}},\ \bibinfo {pages} {032120}
  (\bibinfo {year} {2016})}\BibitemShut {NoStop}%
\bibitem [{\citenamefont {Erker}\ \emph {et~al.}(2017)\citenamefont {Erker},
  \citenamefont {Mitchison}, \citenamefont {Silva}, \citenamefont {Woods},
  \citenamefont {Brunner},\ and\ \citenamefont {Huber}}]{Erker2017}%
  \BibitemOpen
  \bibfield  {author} {\bibinfo {author} {\bibfnamefont {P.}~\bibnamefont
  {Erker}}, \bibinfo {author} {\bibfnamefont {M.~T.}\ \bibnamefont
  {Mitchison}}, \bibinfo {author} {\bibfnamefont {R.}~\bibnamefont {Silva}},
  \bibinfo {author} {\bibfnamefont {M.~P.}\ \bibnamefont {Woods}}, \bibinfo
  {author} {\bibfnamefont {N.}~\bibnamefont {Brunner}}, \ and\ \bibinfo
  {author} {\bibfnamefont {M.}~\bibnamefont {Huber}},\ }\enquote {\bibinfo
  {title} {Autonomous Quantum Clocks: Does Thermodynamics Limit Our Ability to
  Measure Time?}}\ \href {\doibase 10.1103/PhysRevX.7.031022} {\bibfield
  {journal} {\bibinfo  {journal} {Phys. Rev. X}\ }\textbf {\bibinfo {volume}
  {7}},\ \bibinfo {pages} {031022} (\bibinfo {year} {2017})}\BibitemShut
  {NoStop}%
\bibitem [{\citenamefont {Seah}\ \emph {et~al.}(2018)\citenamefont {Seah},
  \citenamefont {Nimmrichter},\ and\ \citenamefont {Scarani}}]{Seah2018}%
  \BibitemOpen
  \bibfield  {author} {\bibinfo {author} {\bibfnamefont {S.}~\bibnamefont
  {Seah}}, \bibinfo {author} {\bibfnamefont {S.}~\bibnamefont {Nimmrichter}}, \
  and\ \bibinfo {author} {\bibfnamefont {V.}~\bibnamefont {Scarani}},\
  }\enquote {\bibinfo {title} {Refrigeration beyond weak internal coupling},}\
  \href {\doibase 10.1103/PhysRevE.98.012131} {\bibfield  {journal} {\bibinfo
  {journal} {Phys. Rev. E}\ }\textbf {\bibinfo {volume} {98}},\ \bibinfo
  {pages} {012131} (\bibinfo {year} {2018})}\BibitemShut {NoStop}%
\bibitem [{\citenamefont {Manzano}\ \emph {et~al.}(2019)\citenamefont
  {Manzano}, \citenamefont {Silva},\ and\ \citenamefont
  {Parrondo}}]{Manzano2019}%
  \BibitemOpen
  \bibfield  {author} {\bibinfo {author} {\bibfnamefont {G.}~\bibnamefont
  {Manzano}}, \bibinfo {author} {\bibfnamefont {R.}~\bibnamefont {Silva}}, \
  and\ \bibinfo {author} {\bibfnamefont {J.~M.~R.}\ \bibnamefont {Parrondo}},\
  }\enquote {\bibinfo {title} {Autonomous thermal machine for amplification and
  control of energetic coherence},}\ \href {\doibase
  10.1103/PhysRevE.99.042135} {\bibfield  {journal} {\bibinfo  {journal} {Phys.
  Rev. E}\ }\textbf {\bibinfo {volume} {99}},\ \bibinfo {pages} {042135}
  (\bibinfo {year} {2019})}\BibitemShut {NoStop}%
\bibitem [{\citenamefont {Man}\ and\ \citenamefont
  {Xia}(2017)}]{man2017smallest}%
  \BibitemOpen
  \bibfield  {author} {\bibinfo {author} {\bibfnamefont {Z.-X.}\ \bibnamefont
  {Man}}\ and\ \bibinfo {author} {\bibfnamefont {Y.-J.}\ \bibnamefont {Xia}},\
  }\enquote {\bibinfo {title} {Smallest quantum thermal machine: The effect of
  strong coupling and distributed thermal tasks},}\ \href {\doibase
  10.1103/PhysRevE.96.012122} {\bibfield  {journal} {\bibinfo  {journal} {Phys.
  Rev. E}\ }\textbf {\bibinfo {volume} {96}},\ \bibinfo {pages} {012122}
  (\bibinfo {year} {2017})}\BibitemShut {NoStop}%
\bibitem [{\citenamefont {Chen}\ \emph {et~al.}(2017)\citenamefont {Chen},
  \citenamefont {Chen},\ and\ \citenamefont {Chen}}]{chen2017thermodynamic}%
  \BibitemOpen
  \bibfield  {author} {\bibinfo {author} {\bibfnamefont {H.-B.}\ \bibnamefont
  {Chen}}, \bibinfo {author} {\bibfnamefont {G.-Y.}\ \bibnamefont {Chen}}, \
  and\ \bibinfo {author} {\bibfnamefont {Y.-N.}\ \bibnamefont {Chen}},\
  }\enquote {\bibinfo {title} {Thermodynamic description of non-Markovian
  information flux of nonequilibrium open quantum systems},}\ \href {\doibase
  10.1103/PhysRevA.96.062114} {\bibfield  {journal} {\bibinfo  {journal} {Phys.
  Rev. A}\ }\textbf {\bibinfo {volume} {96}},\ \bibinfo {pages} {062114}
  (\bibinfo {year} {2017})}\BibitemShut {NoStop}%
\bibitem [{\citenamefont {Clivaz}\ \emph {et~al.}(2019)\citenamefont {Clivaz},
  \citenamefont {Silva}, \citenamefont {Haack}, \citenamefont {Brask},
  \citenamefont {Brunner},\ and\ \citenamefont {Huber}}]{clivaz2019unifying}%
  \BibitemOpen
  \bibfield  {author} {\bibinfo {author} {\bibfnamefont {F.}~\bibnamefont
  {Clivaz}}, \bibinfo {author} {\bibfnamefont {R.}~\bibnamefont {Silva}},
  \bibinfo {author} {\bibfnamefont {G.}~\bibnamefont {Haack}}, \bibinfo
  {author} {\bibfnamefont {J.~B.}\ \bibnamefont {Brask}}, \bibinfo {author}
  {\bibfnamefont {N.}~\bibnamefont {Brunner}}, \ and\ \bibinfo {author}
  {\bibfnamefont {M.}~\bibnamefont {Huber}},\ }\enquote {\bibinfo {title}
  {Unifying paradigms of quantum refrigeration: Fundamental limits of cooling
  and associated work costs},}\ \href {\doibase 10.1103/PhysRevE.100.042130}
  {\bibfield  {journal} {\bibinfo  {journal} {Phys. Rev. E}\ }\textbf {\bibinfo
  {volume} {100}},\ \bibinfo {pages} {042130} (\bibinfo {year}
  {2019})}\BibitemShut {NoStop}%
\bibitem [{\citenamefont {Scovil}\ and\ \citenamefont
  {Schulz-DuBois}(1959)}]{scovil1959three}%
  \BibitemOpen
  \bibfield  {author} {\bibinfo {author} {\bibfnamefont {H.~E.~D.}\
  \bibnamefont {Scovil}}\ and\ \bibinfo {author} {\bibfnamefont {E.~O.}\
  \bibnamefont {Schulz-DuBois}},\ }\enquote {\bibinfo {title} {Three-Level
  Masers as Heat Engines},}\ \href {\doibase 10.1103/PhysRevLett.2.262}
  {\bibfield  {journal} {\bibinfo  {journal} {Phys. Rev. Lett.}\ }\textbf
  {\bibinfo {volume} {2}},\ \bibinfo {pages} {262} (\bibinfo {year}
  {1959})}\BibitemShut {NoStop}%
\bibitem [{\citenamefont {Mitchison}\ \emph {et~al.}(2015)\citenamefont
  {Mitchison}, \citenamefont {Woods}, \citenamefont {Prior},\ and\
  \citenamefont {Huber}}]{Mitchison2015}%
  \BibitemOpen
  \bibfield  {author} {\bibinfo {author} {\bibfnamefont {M.~T.}\ \bibnamefont
  {Mitchison}}, \bibinfo {author} {\bibfnamefont {M.~P.}\ \bibnamefont
  {Woods}}, \bibinfo {author} {\bibfnamefont {J.}~\bibnamefont {Prior}}, \ and\
  \bibinfo {author} {\bibfnamefont {M.}~\bibnamefont {Huber}},\ }\enquote
  {\bibinfo {title} {Coherence-assisted single-shot cooling by quantum
  absorption refrigerators},}\ \href {\doibase 10.1088/1367-2630/17/11/115013}
  {\bibfield  {journal} {\bibinfo  {journal} {New J. Phys.}\ }\textbf {\bibinfo
  {volume} {17}},\ \bibinfo {pages} {115013} (\bibinfo {year}
  {2015})}\BibitemShut {NoStop}%
\bibitem [{\citenamefont {Brask}\ and\ \citenamefont
  {Brunner}(2015)}]{Brask2015}%
  \BibitemOpen
  \bibfield  {author} {\bibinfo {author} {\bibfnamefont {J.~B.}\ \bibnamefont
  {Brask}}\ and\ \bibinfo {author} {\bibfnamefont {N.}~\bibnamefont
  {Brunner}},\ }\enquote {\bibinfo {title} {Small quantum absorption
  refrigerator in the transient regime: Time scales, enhanced cooling, and
  entanglement},}\ \href {\doibase 10.1103/PhysRevE.92.062101} {\bibfield
  {journal} {\bibinfo  {journal} {Phys. Rev. E}\ }\textbf {\bibinfo {volume}
  {92}},\ \bibinfo {pages} {062101} (\bibinfo {year} {2015})}\BibitemShut
  {NoStop}%
\bibitem [{\citenamefont {Lindblad}(1976)}]{lindblad1976generators}%
  \BibitemOpen
  \bibfield  {author} {\bibinfo {author} {\bibfnamefont {G.}~\bibnamefont
  {Lindblad}},\ }\enquote {\bibinfo {title} {On the generators of quantum
  dynamical semigroups},}\ \href {\doibase 10.1007/BF01608499} {\bibfield
  {journal} {\bibinfo  {journal} {Commun. Math. Phys.}\ }\textbf {\bibinfo
  {volume} {48}},\ \bibinfo {pages} {119} (\bibinfo {year} {1976})}\BibitemShut
  {NoStop}%
\bibitem [{\citenamefont {Gorini}\ \emph {et~al.}(1976)\citenamefont {Gorini},
  \citenamefont {Kossakowski},\ and\ \citenamefont
  {Sudarshan}}]{gorini1976completely}%
  \BibitemOpen
  \bibfield  {author} {\bibinfo {author} {\bibfnamefont {V.}~\bibnamefont
  {Gorini}}, \bibinfo {author} {\bibfnamefont {A.}~\bibnamefont {Kossakowski}},
  \ and\ \bibinfo {author} {\bibfnamefont {E.~C.~G.}\ \bibnamefont
  {Sudarshan}},\ }\enquote {\bibinfo {title} {Completely positive dynamical
  semigroups of N-level systems},}\ \href {\doibase 10.1063/1.522979}
  {\bibfield  {journal} {\bibinfo  {journal} {J. Math. Phys.}\ }\textbf
  {\bibinfo {volume} {17}},\ \bibinfo {pages} {821} (\bibinfo {year}
  {1976})}\BibitemShut {NoStop}%
\bibitem [{\citenamefont {Breuer}\ and\ \citenamefont
  {Petruccione}(2002)}]{breuerbook}%
  \BibitemOpen
  \bibfield  {author} {\bibinfo {author} {\bibfnamefont {H.-P.}\ \bibnamefont
  {Breuer}}\ and\ \bibinfo {author} {\bibfnamefont {F.}~\bibnamefont
  {Petruccione}},\ }\href@noop {} {\emph {\bibinfo {title} {The Theory of Open
  Quantum Systems}}}\ (\bibinfo  {publisher} {Oxford University Press},\
  \bibinfo {address} {Oxford},\ \bibinfo {year} {2002})\BibitemShut {NoStop}%
\bibitem [{\citenamefont {Tavakoli}\ \emph {et~al.}(2018)\citenamefont
  {Tavakoli}, \citenamefont {Haack}, \citenamefont {Huber}, \citenamefont
  {Brunner},\ and\ \citenamefont {Brask}}]{tavakoli2018heralded}%
  \BibitemOpen
  \bibfield  {author} {\bibinfo {author} {\bibfnamefont {A.}~\bibnamefont
  {Tavakoli}}, \bibinfo {author} {\bibfnamefont {G.}~\bibnamefont {Haack}},
  \bibinfo {author} {\bibfnamefont {M.}~\bibnamefont {Huber}}, \bibinfo
  {author} {\bibfnamefont {N.}~\bibnamefont {Brunner}}, \ and\ \bibinfo
  {author} {\bibfnamefont {J.~B.}\ \bibnamefont {Brask}},\ }\enquote {\bibinfo
  {title} {Heralded generation of maximal entanglement in any dimension via
  incoherent coupling to thermal baths},}\ \href {\doibase
  10.22331/q-2018-06-13-73} {\bibfield  {journal} {\bibinfo  {journal}
  {{Quantum}}\ }\textbf {\bibinfo {volume} {2}},\ \bibinfo {pages} {73}
  (\bibinfo {year} {2018})}\BibitemShut {NoStop}%
\bibitem [{\citenamefont {Kr\"amer}\ \emph {et~al.}(2018)\citenamefont
  {Kr\"amer}, \citenamefont {Plankensteiner}, \citenamefont {Ostermann},\ and\
  \citenamefont {Ritsch}}]{kraemer2018quantumoptics}%
  \BibitemOpen
  \bibfield  {author} {\bibinfo {author} {\bibfnamefont {S.}~\bibnamefont
  {Kr\"amer}}, \bibinfo {author} {\bibfnamefont {D.}~\bibnamefont
  {Plankensteiner}}, \bibinfo {author} {\bibfnamefont {L.}~\bibnamefont
  {Ostermann}}, \ and\ \bibinfo {author} {\bibfnamefont {H.}~\bibnamefont
  {Ritsch}},\ }\enquote {\bibinfo {title} {QuantumOptics.jl: A Julia framework
  for simulating open quantum systems},}\ \href {\doibase
  10.1016/j.cpc.2018.02.004} {\bibfield  {journal} {\bibinfo  {journal}
  {Comput. Phys. Commun.}\ }\textbf {\bibinfo {volume} {227}},\ \bibinfo
  {pages} {109 } (\bibinfo {year} {2018})}\BibitemShut {NoStop}%
\bibitem [{\citenamefont {Mogensen}\ and\ \citenamefont
  {Riseth}(2018)}]{mogensen2018optim}%
  \BibitemOpen
  \bibfield  {author} {\bibinfo {author} {\bibfnamefont {P.~K.}\ \bibnamefont
  {Mogensen}}\ and\ \bibinfo {author} {\bibfnamefont {A.~N.}\ \bibnamefont
  {Riseth}},\ }\enquote {\bibinfo {title} {Optim: A mathematical optimization
  package for {Julia}},}\ \href {\doibase 10.21105/joss.00615} {\bibfield
  {journal} {\bibinfo  {journal} {J. Open Source Softw.}\ }\textbf {\bibinfo
  {volume} {3}},\ \bibinfo {pages} {615} (\bibinfo {year} {2018})}\BibitemShut
  {NoStop}%
\bibitem [{\citenamefont {Boukobza}\ and\ \citenamefont
  {Tannor}(2007)}]{boukobza2007three}%
  \BibitemOpen
  \bibfield  {author} {\bibinfo {author} {\bibfnamefont {E.}~\bibnamefont
  {Boukobza}}\ and\ \bibinfo {author} {\bibfnamefont {D.~J.}\ \bibnamefont
  {Tannor}},\ }\enquote {\bibinfo {title} {Three-Level Systems as Amplifiers
  and Attenuators: A Thermodynamic Analysis},}\ \href {\doibase
  10.1103/PhysRevLett.98.240601} {\bibfield  {journal} {\bibinfo  {journal}
  {Phys. Rev. Lett.}\ }\textbf {\bibinfo {volume} {98}},\ \bibinfo {pages}
  {240601} (\bibinfo {year} {2007})}\BibitemShut {NoStop}%
\bibitem [{\citenamefont {Niedenzu}\ \emph {et~al.}(2019)\citenamefont
  {Niedenzu}, \citenamefont {Huber},\ and\ \citenamefont
  {Boukobza}}]{niedenzu2019concepts}%
  \BibitemOpen
  \bibfield  {author} {\bibinfo {author} {\bibfnamefont {W.}~\bibnamefont
  {Niedenzu}}, \bibinfo {author} {\bibfnamefont {M.}~\bibnamefont {Huber}}, \
  and\ \bibinfo {author} {\bibfnamefont {E.}~\bibnamefont {Boukobza}},\
  }\enquote {\bibinfo {title} {Concepts of work in autonomous quantum heat
  engines},}\ \href {\doibase 10.22331/q-2019-10-14-195} {\bibfield  {journal}
  {\bibinfo  {journal} {{Quantum}}\ }\textbf {\bibinfo {volume} {3}},\ \bibinfo
  {pages} {195} (\bibinfo {year} {2019})}\BibitemShut {NoStop}%
\bibitem [{\citenamefont {Ghosh}\ \emph {et~al.}(2018)\citenamefont {Ghosh},
  \citenamefont {Gelbwaser-Klimovsky}, \citenamefont {Niedenzu}, \citenamefont
  {Lvovsky}, \citenamefont {Mazets}, \citenamefont {Scully},\ and\
  \citenamefont {Kurizki}}]{ghosh2018twolevel}%
  \BibitemOpen
  \bibfield  {author} {\bibinfo {author} {\bibfnamefont {A.}~\bibnamefont
  {Ghosh}}, \bibinfo {author} {\bibfnamefont {D.}~\bibnamefont
  {Gelbwaser-Klimovsky}}, \bibinfo {author} {\bibfnamefont {W.}~\bibnamefont
  {Niedenzu}}, \bibinfo {author} {\bibfnamefont {A.~I.}\ \bibnamefont
  {Lvovsky}}, \bibinfo {author} {\bibfnamefont {I.}~\bibnamefont {Mazets}},
  \bibinfo {author} {\bibfnamefont {M.~O.}\ \bibnamefont {Scully}}, \ and\
  \bibinfo {author} {\bibfnamefont {G.}~\bibnamefont {Kurizki}},\ }\enquote
  {\bibinfo {title} {Two-level masers as heat-to-work converters},}\ \href
  {\doibase 10.1073/pnas.1805354115} {\bibfield  {journal} {\bibinfo  {journal}
  {PNAS}\ }\textbf {\bibinfo {volume} {115}},\ \bibinfo {pages} {9941}
  (\bibinfo {year} {2018})}\BibitemShut {NoStop}%
\bibitem [{\citenamefont {Boukobza}\ and\ \citenamefont
  {Tannor}(2008)}]{Boukobza2008}%
  \BibitemOpen
  \bibfield  {author} {\bibinfo {author} {\bibfnamefont {E.}~\bibnamefont
  {Boukobza}}\ and\ \bibinfo {author} {\bibfnamefont {D.~J.}\ \bibnamefont
  {Tannor}},\ }\enquote {\bibinfo {title} {Thermodynamic analysis of quantum
  light purification},}\ \href {\doibase 10.1103/PhysRevA.78.013825} {\bibfield
   {journal} {\bibinfo  {journal} {Phys. Rev. A}\ }\textbf {\bibinfo {volume}
  {78}},\ \bibinfo {pages} {013825} (\bibinfo {year} {2008})}\BibitemShut
  {NoStop}%
\bibitem [{\citenamefont {Youssef}\ \emph {et~al.}(2009)\citenamefont
  {Youssef}, \citenamefont {Mahler},\ and\ \citenamefont
  {Obada}}]{Youssef2009}%
  \BibitemOpen
  \bibfield  {author} {\bibinfo {author} {\bibfnamefont {M.}~\bibnamefont
  {Youssef}}, \bibinfo {author} {\bibfnamefont {G.}~\bibnamefont {Mahler}}, \
  and\ \bibinfo {author} {\bibfnamefont {A.-S.~F.}\ \bibnamefont {Obada}},\
  }\enquote {\bibinfo {title} {Quantum optical thermodynamic machines: Lasing
  as relaxation},}\ \href {\doibase 10.1103/PhysRevE.80.061129} {\bibfield
  {journal} {\bibinfo  {journal} {Phys. Rev. E}\ }\textbf {\bibinfo {volume}
  {80}},\ \bibinfo {pages} {061129} (\bibinfo {year} {2009})}\BibitemShut
  {NoStop}%
\bibitem [{\citenamefont {Boukobza}\ and\ \citenamefont
  {Tannor}(2006)}]{boukobza2006thermodynamic}%
  \BibitemOpen
  \bibfield  {author} {\bibinfo {author} {\bibfnamefont {E.}~\bibnamefont
  {Boukobza}}\ and\ \bibinfo {author} {\bibfnamefont {D.~J.}\ \bibnamefont
  {Tannor}},\ }\enquote {\bibinfo {title} {Thermodynamic analysis of quantum
  light amplification},}\ \href {\doibase 10.1103/PhysRevA.74.063822}
  {\bibfield  {journal} {\bibinfo  {journal} {Phys. Rev. A}\ }\textbf {\bibinfo
  {volume} {74}},\ \bibinfo {pages} {063822} (\bibinfo {year}
  {2006})}\BibitemShut {NoStop}%
\bibitem [{\citenamefont {Liesen}\ and\ \citenamefont
  {Mehrmann}(2015)}]{liesen2002linear}%
  \BibitemOpen
  \bibfield  {author} {\bibinfo {author} {\bibfnamefont {J.}~\bibnamefont
  {Liesen}}\ and\ \bibinfo {author} {\bibfnamefont {V.}~\bibnamefont
  {Mehrmann}},\ }\href {\doibase https://doi.org/10.1007/978-3-319-24346-7}
  {\emph {\bibinfo {title} {Linear Algebra}}}\ (\bibinfo  {publisher}
  {Springer, Cham},\ \bibinfo {year} {2015})\BibitemShut {NoStop}%
\bibitem [{\citenamefont {Rignon-Bret}\ \emph {et~al.}(2021)\citenamefont
  {Rignon-Bret}, \citenamefont {Guarnieri}, \citenamefont {Goold},\ and\
  \citenamefont {Mitchison}}]{rignon2021thermodynamics}%
  \BibitemOpen
  \bibfield  {author} {\bibinfo {author} {\bibfnamefont {A.}~\bibnamefont
  {Rignon-Bret}}, \bibinfo {author} {\bibfnamefont {G.}~\bibnamefont
  {Guarnieri}}, \bibinfo {author} {\bibfnamefont {J.}~\bibnamefont {Goold}}, \
  and\ \bibinfo {author} {\bibfnamefont {M.~T.}\ \bibnamefont {Mitchison}},\
  }\enquote {\bibinfo {title} {Thermodynamics of precision in quantum
  nanomachines},}\ \href {\doibase 10.1103/PhysRevE.103.012133} {\bibfield
  {journal} {\bibinfo  {journal} {Phys. Rev. E}\ }\textbf {\bibinfo {volume}
  {103}},\ \bibinfo {pages} {012133} (\bibinfo {year} {2021})}\BibitemShut
  {NoStop}%
\end{thebibliography}%

\appendix

\section{Reset master equation}

\label{app:RME}

To justify the form of the RME~{\eqref{eq:master_1vq_en}}, here we loosely follow Appendix~A in Ref.~{\cite{Linden2010}} and the main text in Ref.~{\cite{skrzypczyk2011smallest}}. The RME is a simple model that describes a probabilistic swapping with a thermal state whose temperature corresponds to that of the environment instead of illustrating some physical interaction with the environment. Consider a system of three qubits coupled to baths depicted in Fig.~{\ref{fig:qubit_vq_Ten}}. The RME expresses the thermalisation processes for the qubits such that each qubit is ``reset'' to a thermal state at the temperature of its bath with probability density $p_i$ per time $\delta t$. Namely, to first order in $\delta t$, the density matrix at $t+\delta t$ is provided by
\begin{align}
    \rho_{\mathrm{tot}}(t+\delta t)
    &=
    \left(
    1-\sum_{i\in\{\mathrm{en},1,2\}}p_i\delta t
    \right) \rho_{\mathrm{tot}}
    \nonumber\\
    &\quad
    +\sum_{i\in\{\mathrm{en},1,2\}}
    p_i \delta t
    \left(
    \tau_i \otimes \mathrm{Tr}_i [\rho_{\mathrm{tot}}]
    \right)
    -i\delta t
    \left[
    H,\rho_{\mathrm{tot}}
    \right]
    \nonumber\\
    &=
    \rho_{\mathrm{tot}}
    +
    \sum_{i\in\{\mathrm{en},1,2\}}
    p_i \delta t
    \left(
    \tau_i \otimes \mathrm{Tr}_i [\rho_{\mathrm{tot}}] - \rho_{\mathrm{tot}}
    \right)
    \nonumber\\
    &\quad
    -i\delta t
    \left[
    H,\rho_{\mathrm{tot}}
    \right],
\end{align}
which leads to Eq.~{\eqref{eq:master_1vq_en}}.

\section{Partial traces for a single system in our notation}

\label{app:partialtrace}

While partial traces and tensor products are usually defined for composite systems, we have used a slight modification of this notation to simplify our treatment of higher-dimensional systems in our context. Here, we further explain the notation $\mathrm{Tr}_{\mathrm{A,B,C}}$ introduced in Eq.~{\eqref{eq:master_ABC}} of Sec.~{\ref{sec_qutrit}}. Let us remind the reader that we consider a three-level system where all of the level pairs are in contact with different baths (see the right hand side of Fig.~{\ref{fig:qutrit_6qubits}}). Here, A, B, and C stand for the level pair of $\ket{0}$ and $\ket{1}$, the one of $\ket{0}$ and $\ket{2}$, and the one of $\ket{1}$ and $\ket{2}$, respectively. First of all, a point is that off-diagonal terms of the density matrix vanish. Therefore, the density matrix of any state in the target qutrit is given by
\begin{align}
    \rho
    =
    \begin{pmatrix}
    p_0 & 0 & 0 \\
    0 & p_1 & 0 \\
    0 & 0 & p_2 
    \end{pmatrix}.
\end{align}
Here, we define $\mathrm{Tr}_{\mathrm{A}}[\rho]$ as
\begin{align}
    \mathrm{Tr}_{\mathrm{A}}[\rho] 
    =
    \begin{pmatrix}
    p_0+p_1 & 0 \\
    0 & p_2
    \end{pmatrix}.
\end{align}
A tensor product of this partially-traced-out state $\mathrm{Tr}_{\mathrm{A}}[\rho]$ and a thermal state $\tau_{\mathrm{A}}=(\tau_{\mathrm{A}}^{\mathrm{g}},\tau_{\mathrm{A}}^{\mathrm{e}})^{\mathrm{T}}$ appears in the summation of Eq.~{\eqref{eq:master_ABC}} and means a state where the population ratio between the levels $\ket{0}$ and $\ket{1}$ is $\exp[-(\omega_1-\omega_0)/T_{\mathrm{vA}}]$. Accordingly, we write this tensor product as
\begin{align}
    \tau_{\mathrm{A}}\otimes\mathrm{Tr}_{\mathrm{A}}[\rho]
    =
    \begin{pmatrix}
    (p_0+p_1)\tau_{\mathrm{A}}^{\mathrm{g}} & 0 & 0 \\
    0 & (p_0+p_1)\tau_{\mathrm{A}}^{\mathrm{e}} & 0 \\
    0 & 0 & p_2 
    \end{pmatrix}
\end{align}
due to $\tau_{\mathrm{A}}^{\mathrm{e}}/\tau_{\mathrm{A}}^{\mathrm{g}}=\exp[-(\omega_1-\omega_0)/T_{\mathrm{vA}}]$. If one applies the trace-out $\mathrm{Tr}_{\mathrm{A}}$ to $\tau_{\mathrm{A}}\otimes\mathrm{Tr}_{\mathrm{A}}[\rho]$, it should be $\mathrm{Tr}_{\mathrm{A}}[\rho]$, and one can verify this as 
\begin{align}
    \mathrm{Tr}_{\mathrm{A}}\left[
    \tau_{\mathrm{A}}\otimes\mathrm{Tr}_{\mathrm{A}}[\rho]
    \right]
    &=
    \begin{pmatrix}
    (p_0+p_1)\tau_{\mathrm{A}}^{\mathrm{g}}+(p_0+p_1)\tau_{\mathrm{A}}^{\mathrm{e}} & 0 \\
    0 & p_2
    \end{pmatrix}
    \nonumber\\
    &
    \begin{pmatrix}
    p_0+p_1 & 0 \\
    0 & p_2 
    \end{pmatrix}
    \nonumber\\
    &
    \mathrm{Tr}_{\mathrm{A}}[\rho].
\end{align}
due to $\tau_{\mathrm{A}}^{\mathrm{g}}+\tau_{\mathrm{A}}^{\mathrm{e}}=1$.

\section{Steady-state solution of effRME for $n$-level target system}

\label{app:derivation_steady_n}

We discuss the steady-state solution for an effRME in a multi-level system with some two-qubit machines coupled. For simplicity, let us adhere to cases where every pair of levels in the target is coupled to one machine. In these cases, for $n$-level systems the number of the couplings is $\binom{n}{2}=n(n-1)/2$.

We generalise the effRME to $n$-level target systems. For distinct representation, let us introduce a different notation of coupling strength from that in Fig.~\ref{fig:qutrit_6qubits}. We write $q_{k,l}$ as the thermalisation rate of the $k$th and $l$th levels ($k<l$), where the indices A, B, C in Fig.~\ref{fig:qutrit_6qubits} are associated with $q_{0,1}$, $q_{0,2}$, and $q_{1,2}$, respectively. The effRME for $n$-level target system is written as
\begin{equation} \label{eq:master_kl}
    \frac{\partial \rho}{\partial t}
    =
    \sum_{l=1}^{n-1}
    \sum_{k=0}^{l-1}
    q_{k,l} 
    \left(
    \tau_{k,l} \otimes \mathrm{Tr}_{k,l} [\rho]
    - \rho
    \right),
\end{equation}
where $\tau_{k,l}$ is a thermal state at the virtual temperature associated with the $k$th and $l$th levels, and $\mathrm{Tr}_{k,l}$ represents tracing out the space of the $k$th and $l$th levels. We ignore off-diagonal terms in the density matrix since in this model coherence cannot be generated. Then, this equation can be simplified as 
\begin{equation} \label{eq:master_n}
    \frac{\partial \rho}{\partial t}
    =
    \sum_{l=1}^{n-1}
    \sum_{k=0}^{l-1}
    q_{k,l} 
    \left(
    -\tau_{k,l}^{\mathrm{e}} \rho^{(k)} + \tau_{k,l}^{\mathrm{g}} \rho^{(l)}
    \right)
    \left(
    \ketbra{k}{k} - \ketbra{l}{l}
    \right)
    ,
\end{equation}
with $\rho^{(k)}=\brakket{k}{\rho}{k}$. To obtain the steady state, we solve $\partial\rho/\partial t =0$, i.e. 
\begin{equation} \label{eq:master_Cij}
    \sum_{l=1}^{n-1}
    \sum_{k=0}^{l-1}
    C_{k,l}
    \left(
    \ketbra{k}{k}
    -
    \ketbra{l}{l}
    \right)
    =
    0, 
\end{equation}
where $C_{k,l}=q_{k,l} \left(- \tau_{k,l}^{\mathrm{e}} \rho^{(k)} + \tau_{k,l}^{\mathrm{g}} \rho^{(l)} \right)$.

First, let us separate the equation into two terms as 
\begin{equation} \label{eq:master_n_two}
    \sum_{l=1}^{n-1}
    \sum_{k=0}^{l-1}
    C_{k,l}
    \ketbra{k}{k}
    -
    \sum_{l=1}^{n-1}
    \sum_{k=0}^{l-1}
    C_{k,l}
    \ketbra{l}{l}
    =
    0.
\end{equation}
The first term can be written in a different way, 
\begin{align}
    \sum_{l=1}^{n-1}
    \sum_{k=0}^{l-1}
    C_{k,l}
    \ketbra{k}{k}
    &=
    \sum_{l=1}^{n-1} C_{0,l}
    \ketbra{0}{0}
    +
    \sum_{l=2}^{n-1} C_{1,l}
    \ketbra{1}{1}
    \nonumber\\
    &\quad 
    +
    \cdots
    +
    \sum_{l=n-1}^{n-1} C_{n-2,l}
    \ketbra{n-2}{n-2}
    \nonumber\\
    &=
    \sum_{s=0}^{n-2}
    \sum_{l=s+1}^{n-1} C_{s,l}
    \ketbra{s}{s}
    \nonumber\\
    &=
    \sum_{l=1}^{n-1} C_{0,l}
    \ketbra{0}{0}
    +
    \sum_{s=1}^{n-2}
    \sum_{l=s+1}^{n-1} C_{s,l}
    \ketbra{s}{s},
\end{align}
and the second term can be written as 
\begin{equation}
    \sum_{l=1}^{n-1}
    \sum_{k=0}^{l-1}
    C_{k,l}
    \ketbra{l}{l}
    =
    \sum_{l=1}^{n-2}
    \sum_{k=0}^{l-1}
    C_{k,l}
    \ketbra{l}{l}
    +
    \sum_{k=0}^{n-2}
    C_{k,n-1}
    \ketbra{n-1}{n-1}.
\end{equation}
Thus, the left hand side (l.h.s) of Eq.~\eqref{eq:master_n_two} is rewritten as
\begin{align}
    \text{l.h.s of Eq.~\eqref{eq:master_n_two}}
    &=
    \sum_{l=1}^{n-1} C_{0,l}
    \ketbra{0}{0}
    -
    \sum_{k=0}^{n-2}
    C_{k,n-1}
    \ketbra{n-1}{n-1}
    \nonumber
    \\
    &\quad
    +
    \sum_{s=1}^{n-2}
    \sum_{l=s+1}^{n-1} C_{s,l}
    \ketbra{s}{s}
    -
    \sum_{l=1}^{n-2}
    \sum_{k=0}^{l-1}
    C_{k,l}
    \ketbra{l}{l}
    \nonumber
    \\
    &=
    \sum_{l=1}^{n-1} C_{0,l}
    \ketbra{0}{0}
    -
    \sum_{k=0}^{n-2}
    C_{k,n-1}
    \ketbra{n-1}{n-1}
    \nonumber
    \\
    &\quad
    +
    \sum_{s=1}^{n-2}
    \left(
    \sum_{l=s+1}^{n-1} C_{s,l}
    -
    \sum_{k=0}^{s-1}
    C_{k,s}
    \right)
    \ketbra{s}{s}.
\end{align}
Since each of the terms in Eq.~\eqref{eq:master_n_two} is zero, we can obtain $n$ equations such as
\begin{subequations}
\begin{align}
    \sum_{l=s+1}^{n-1} C_{s,l}
    -
    \sum_{k=0}^{s-1}
    C_{k,s}
    &=
    0,
    \,\,
    \text{for}\,\left\{
    1\leq k \leq n-2:\forall k\in\mathbb{Z}
    \right\},
    \\
    \sum_{l=1}^{n-1} C_{0,l}
    &=
    0,
    \\
    \sum_{k=0}^{n-2} C_{k,n-1}
    &=
    0.
\end{align}
\end{subequations}
The above $n$ equations can be written in a matrix form as 
\begin{equation} \label{eq:Mrho0}
    \mathbf{M}_{n} \vec{\rho}_{\mathrm{ss}}
    =
    \vec{0}
\end{equation}
where $\vec{\rho}_{\mathrm{ss}}=(\rho^{(0)}_{\mathrm{ss}},\rho^{(1)}_{\mathrm{ss}},\ldots,\rho^{(n-1)}_{\mathrm{ss}})^{T}$ and $\mathbf{M}_n$ is an $n\times n$ matrix given by 
\begin{widetext}
\begin{equation} \label{eq:matrixM0}
    \mathbf{M}
    =
    \begin{pmatrix}
    M_{0,0} & q_{0,1} \tau_{0,1}^{\mathrm{g}} & q_{0,2} \tau_{0,2}^{\mathrm{g}} & \cdots & q_{0,n-2} \tau_{0,n-2}^{\mathrm{g}} & q_{0,n-1} \tau_{0,n-1}^{\mathrm{g}}
    \\
    q_{0,1} \tau_{0,1}^{\mathrm{e}} & M_{1,1} & q_{1,2} \tau_{1,2}^{\mathrm{g}} & \cdots & q_{1,n-2} \tau_{1,n-2}^{\mathrm{g}} & q_{1,n-1} \tau_{1,n-1}^{\mathrm{g}}
    \\
    q_{0,2} \tau_{0,2}^{\mathrm{e}} & q_{1,2} \tau_{1,2}^{\mathrm{e}} & \ddots & \ddots & \vdots & \vdots
    \\
    \vdots & \vdots & {} & {} & \vdots & \vdots
    \\
    \vdots & \vdots & \ddots & \ddots & q_{n-3,n-2} \tau_{n-3,n-2}^{\mathrm{g}} & q_{n-3,n-1} \tau_{n-3,n-1}^{\mathrm{g}}
    \\
    q_{0,n-2} \tau_{0,n-2}^{\mathrm{e}} & q_{1,n-2} \tau_{1,n-2}^{\mathrm{e}} & \cdots & q_{n-3,n-2} \tau_{n-3,n-2}^{\mathrm{e}} & M_{n-2,n-2} & q_{n-2,n-1} \tau_{n-2,n-1}^{\mathrm{g}}
    \\
    q_{0,n-1} \tau_{0,n-1}^{\mathrm{e}} & q_{1,n-1} \tau_{1,n-1}^{\mathrm{e}} & \cdots & q_{n-3,n-1} \tau_{n-3,n-1}^{\mathrm{e}} &  q_{n-2,n-1} \tau_{n-2,n-1}^{\mathrm{e}} & M_{n-1,n-1}
    \end{pmatrix}\; .
\end{equation}
\end{widetext}
The diagonal terms are given by
\begin{subequations} \label{eq:diagM}
\begin{align}
    M_{0,0} \label{eq:diagM00}
    &=-\sum_{s=1}^{n-1} q_{0,s} \tau_{0,s}^{\mathrm{e}}\;, 
    \\
    M_{k,k} \label{eq:diagMkk}
    &=
    -\left(\sum_{s=0}^{k-1}q_{s,k} \tau_{s,k}^{\mathrm{g}} + \sum_{s=k+1}^{n-1} q_{k,s} \tau_{k,s}^{\mathrm{e}} \right)\;,
    \\
    M_{n-1,n-1} \label{eq:diagMn-1n-1}
    &=-\sum_{s=0}^{n-2} q_{s,n-1} \tau_{s,n-1}^{\mathrm{g}}\;
\end{align}
\end{subequations}
for $1\leq k \leq n-2$. 

Here, we add the normalisation constraint, $\mathrm{Tr}[\rho_{\mathrm{ss}}]=1$, into this simultaneous equation~\eqref{eq:Mrho0}, and hence the total number of equations involved in the simultaneous equation is $(n+1)$. However, the number of the variables in $\vec{\rho}_{\mathrm{ss}}$ is $n$. This indicates that there is one excess equation in the simultaneous equation. 
In fact, any equation written inside {Eq.~\eqref{eq:Mrho0}} is dependent on other equations, (i.e. can be constructed from the rest of the equations). 
For example, the equation described by the first row in the matrix $\mathbf{M}$ is reproduced by taking a sum of the equations given by all other rows due to Eqs.~\eqref{eq:diagM} and multiplying it by a minus sign. Therefore, the removal of the first row from the matrix $\mathbf{M}$ poses no problem for solving the simultaneous equation. 
We remove the first row and then add the normalisation constraint $\mathrm{Tr}[\rho_{\mathrm{ss}}]=1$ as follows
\begin{widetext}
\begin{align}
    \mathbf{M}_{n} \vec{\rho}_{\mathrm{ss}}
    &=
    \begin{pmatrix}
    M_{0,0} & q_{0,1} \tau_{0,1}^{\mathrm{g}} & q_{0,2} \tau_{0,2}^{\mathrm{g}} & \cdots & q_{0,n-2} \tau_{0,n-2}^{\mathrm{g}} & q_{0,n-1} \tau_{0,n-1}^{\mathrm{g}}
    \\
    q_{0,1} \tau_{0,1}^{\mathrm{e}} & M_{1,1} & q_{1,2} \tau_{1,2}^{\mathrm{g}} & \cdots & q_{1,n-2} \tau_{1,n-2}^{\mathrm{g}} & q_{1,n-1} \tau_{1,n-1}^{\mathrm{g}}
    \\
    q_{0,2} \tau_{0,2}^{\mathrm{e}} & q_{1,2} \tau_{1,2}^{\mathrm{e}} & \ddots & \ddots & \vdots & \vdots
    \\
    \vdots & \vdots & {} & {} & \vdots & \vdots
    \\
    \vdots & \vdots & \ddots & \ddots & q_{n-3,n-2} \tau_{n-3,n-2}^{\mathrm{g}} & q_{n-3,n-1} \tau_{n-3,n-1}^{\mathrm{g}}
    \\
    q_{0,n-2} \tau_{0,n-2}^{\mathrm{e}} & q_{1,n-2} \tau_{1,n-2}^{\mathrm{e}} & \cdots & q_{n-3,n-2} \tau_{n-3,n-2}^{\mathrm{e}} & M_{n-2,n-2} & q_{n-2,n-1} \tau_{n-2,n-1}^{\mathrm{g}}
    \\
    q_{0,n-1} \tau_{0,n-1}^{\mathrm{e}} & q_{1,n-1} \tau_{1,n-1}^{\mathrm{e}} & \cdots & q_{n-3,n-1} \tau_{n-3,n-1}^{\mathrm{e}} &  q_{n-2,n-1} \tau_{n-2,n-1}^{\mathrm{e}} & M_{n-1,n-1}
    \end{pmatrix}
    \begin{pmatrix}
    \rho_{\mathrm{ss}}^{(0)} \\ 
    \rho_{\mathrm{ss}}^{(1)} \\
    \vdots \\
    \vdots \\
    \vdots \\
    \vdots \\
    \rho_{\mathrm{ss}}^{(n-1)}
    \end{pmatrix}
    \nonumber\\
    &\to
    \begin{pmatrix}
    0 & 0 & 0 & \cdots & 0 & 0
    \\
    q_{0,1} \tau_{0,1}^{\mathrm{e}} & M_{1,1} & q_{1,2} \tau_{1,2}^{\mathrm{g}} & \cdots & q_{1,n-2} \tau_{1,n-2}^{\mathrm{g}} & q_{1,n-1} \tau_{1,n-1}^{\mathrm{g}}
    \\
    q_{0,2} \tau_{0,2}^{\mathrm{e}} & q_{1,2} \tau_{1,2}^{\mathrm{e}} & \ddots & \ddots & \vdots & \vdots
    \\
    \vdots & \vdots & {} & {} & \vdots & \vdots
    \\
    \vdots & \vdots & \ddots & \ddots & q_{n-3,n-2} \tau_{n-3,n-2}^{\mathrm{g}} & q_{n-3,n-1} \tau_{n-3,n-1}^{\mathrm{g}}
    \\
    q_{0,n-2} \tau_{0,n-2}^{\mathrm{e}} & q_{1,n-2} \tau_{1,n-2}^{\mathrm{e}} & \cdots & q_{n-3,n-2} \tau_{n-3,n-2}^{\mathrm{e}} & M_{n-2,n-2} & q_{n-2,n-1} \tau_{n-2,n-1}^{\mathrm{g}}
    \\
    q_{0,n-1} \tau_{0,n-1}^{\mathrm{e}} & q_{1,n-1} \tau_{1,n-1}^{\mathrm{e}} & \cdots & q_{n-3,n-1} \tau_{n-3,n-1}^{\mathrm{e}} &  q_{n-2,n-1} \tau_{n-2,n-1}^{\mathrm{e}} & M_{n-1,n-1}
    \end{pmatrix}
    \begin{pmatrix}
    \rho_{\mathrm{ss}}^{(0)} \\ 
    \rho_{\mathrm{ss}}^{(1)} \\
    \vdots \\
    \vdots \\
    \vdots \\
    \vdots \\
    \rho_{\mathrm{ss}}^{(n-1)}
    \end{pmatrix}
    \nonumber\\
    &\to
    \begin{pmatrix}
    1 & 1 & 1 & \cdots & 1 & 1
    \\
    q_{0,1} \tau_{0,1}^{\mathrm{e}} & M_{1,1} & q_{1,2} \tau_{1,2}^{\mathrm{g}} & \cdots & q_{1,n-2} \tau_{1,n-2}^{\mathrm{g}} & q_{1,n-1} \tau_{1,n-1}^{\mathrm{g}}
    \\
    q_{0,2} \tau_{0,2}^{\mathrm{e}} & q_{1,2} \tau_{1,2}^{\mathrm{e}} & \ddots & \ddots & \vdots & \vdots
    \\
    \vdots & \vdots & {} & {} & \vdots & \vdots
    \\
    \vdots & \vdots & \ddots & \ddots & q_{n-3,n-2} \tau_{n-3,n-2}^{\mathrm{g}} & q_{n-3,n-1} \tau_{n-3,n-1}^{\mathrm{g}}
    \\
    q_{0,n-2} \tau_{0,n-2}^{\mathrm{e}} & q_{1,n-2} \tau_{1,n-2}^{\mathrm{e}} & \cdots & q_{n-3,n-2} \tau_{n-3,n-2}^{\mathrm{e}} & M_{n-2,n-2} & q_{n-2,n-1} \tau_{n-2,n-1}^{\mathrm{g}}
    \\
    q_{0,n-1} \tau_{0,n-1}^{\mathrm{e}} & q_{1,n-1} \tau_{1,n-1}^{\mathrm{e}} & \cdots & q_{n-3,n-1} \tau_{n-3,n-1}^{\mathrm{e}} &  q_{n-2,n-1} \tau_{n-2,n-1}^{\mathrm{e}} & M_{n-1,n-1}
    \end{pmatrix}
    \begin{pmatrix}
    \rho_{\mathrm{ss}}^{(0)} \\ 
    \rho_{\mathrm{ss}}^{(1)} \\
    \vdots \\
    \vdots \\
    \vdots \\
    \vdots \\
    \rho_{\mathrm{ss}}^{(n-1)}
    \end{pmatrix}\; .
\end{align}
The full simultaneous equation turns to become
\begin{equation} \label{eq:fullequation_Mrho10}
    \mathbf{M}
    \begin{pmatrix}
    \rho_{\mathrm{ss}}^{(0)} \\ 
    \rho_{\mathrm{ss}}^{(1)} \\
    \vdots \\
    \rho_{\mathrm{ss}}^{(n-1)}
    \end{pmatrix}
    =
    \begin{pmatrix}
    1 \\ 
    0 \\
    \vdots \\
    0
    \end{pmatrix}\; ,
\end{equation}
where the matrix $\mathbf{M}$ is now redefined as 
\begin{equation} \label{eq:matrixM}
    \mathbf{M}
    =
    \begin{pmatrix}
    1 & 1 & 1 & \cdots & 1 & 1
    \\
    q_{0,1} \tau_{0,1}^{\mathrm{e}} & M_{1,1} & q_{1,2} \tau_{1,2}^{\mathrm{g}} & \cdots & q_{1,n-2} \tau_{1,n-2}^{\mathrm{g}} & q_{1,n-1} \tau_{1,n-1}^{\mathrm{g}}
    \\
    q_{0,2} \tau_{0,2}^{\mathrm{e}} & q_{1,2} \tau_{1,2}^{\mathrm{e}} & \ddots & \ddots & \vdots & \vdots
    \\
    \vdots & \vdots & {} & {} & \vdots & \vdots
    \\
    \vdots & \vdots & \ddots & \ddots & q_{n-3,n-2} \tau_{n-3,n-2}^{\mathrm{g}} & q_{n-3,n-1} \tau_{n-3,n-1}^{\mathrm{g}}
    \\
    q_{0,n-2} \tau_{0,n-2}^{\mathrm{e}} & q_{1,n-2} \tau_{1,n-2}^{\mathrm{e}} & \cdots & q_{n-3,n-2} \tau_{n-3,n-2}^{\mathrm{e}} & M_{n-2,n-2} & q_{n-2,n-1} \tau_{n-2,n-1}^{\mathrm{g}}
    \\
    q_{0,n-1} \tau_{0,n-1}^{\mathrm{e}} & q_{1,n-1} \tau_{1,n-1}^{\mathrm{e}} & \cdots & q_{n-3,n-1} \tau_{n-3,n-1}^{\mathrm{e}} &  q_{n-2,n-1} \tau_{n-2,n-1}^{\mathrm{e}} & M_{n-1,n-1}
    \end{pmatrix}\;.
\end{equation}
\end{widetext}

Here, let us distinguish the two cases where the matrix $\mathbf{M}$ is invertible and where it is not. In the latter case, the steady-state solution cannot be determined with the conditions we have. However, this issue can be avoided. For example, physically, this is the case where one machine is coupled to the levels $\ket{0}$ and $\ket{1}$ in a three-level system and the population ratios between the levels $\ket{0}$ and $\ket{2}$ and between the levels $\ket{1}$ and $\ket{2}$ are not determined. In this case, the steady state is not unique, and this leads to the nonexistence of inverse matrix of $\mathbf{M}$. If one sees this three-level system as a two-level system composed of the levels $\ket{0}$ and $\ket{1}$, the matrix $\mathbf{M}$ can be rewritten as an invertible matrix.

Assuming that the matrix $\mathbf{M}$ is invertible, the solution $\vec{\rho}_{\mathrm{ss}}$ is obtained as 
\begin{equation}
    \vec{\rho}_{\mathrm{ss}}
    =
    \mathbf{M}^{-1}
    \begin{pmatrix}
    1 \\ 
    0 \\
    \vdots \\
    0
    \end{pmatrix}\; .
\end{equation}
According to Cramer's rule~\cite{liesen2002linear}, the inverse matrix can be written as 
\begin{equation}
    \mathbf{M}^{-1}
    =
    \frac{1
    }{
    \det[\mathbf{M}]
    }
    \mathrm{adj}
    \left[
    \mathbf{M}
    \right], 
\end{equation}
where $\mathrm{adj}[\mathbf{M}]$ is the adjugate of $\mathbf{M}$, given by $\mathrm{adj}[\mathbf{M}]=[\{\Delta_{i,j}\}_{1\leq i,j \leq n}]^{T}$, i.e.
\begin{equation}
    \mathrm{adj}[\mathbf{M}]
    =
    \begin{pmatrix}
    \Delta_{1,1} & \Delta_{2,1} & \cdots & \Delta_{n,1} \\
    \Delta_{1,2} & \Delta_{2,2} & \cdots & \Delta_{n,2} \\
    \vdots & \vdots & \ddots & \vdots \\
    \Delta_{1,n} & \Delta_{2,n} & \cdots & \Delta_{n,n}
    \end{pmatrix}\; .
\end{equation}
Here, $\Delta_{i,j}$ is a set of the cofactors of the matrix $\mathbf{M}$ and defined as 
\begin{align} \label{eq:Deltaij}
    &\Delta_{i,j}
    =
    \nonumber\\
    &
    (-1)^{i+j}
    \begin{vmatrix}
    M_{0,0} & \cdots & 
    M_{0,j-1} & 
    M_{0,j+1} & \cdots & 
    M_{0,n-1} \\
    \vdots & \vdots & \vdots & \vdots & \vdots & \vdots \\
    M_{i-1,0} & \cdots & M_{i-1,j-1} & 
    M_{i-1,j+1} & \cdots & M_{i-1,n-1} \\
    M_{i+1,0} & \cdots & M_{i+1,j-1} & 
    M_{i+1,j+1} & \cdots & M_{i+1,n-1} \\
    \vdots & \vdots & \vdots & \vdots & \vdots & \vdots \\
    M_{n-1,0} & \cdots & M_{n-1,j-1} & 
    M_{n-1,j+1} & \cdots & M_{n-1,n-1}
    \end{vmatrix}\;.
\end{align}
Also, due to the mathematical properties of the determinant, we get $\det[\mathbf{M}]=\sum_{s=1}^{n}\Delta_{1,s}$. As a result, the solution $\vec{\rho}_{\mathrm{ss}}$ is then written as 
\begin{equation} \label{eq:steady_n}
    \vec{\rho}_{\mathrm{ss}}
    =
    \frac{1}{\sum_{s=1}^{n}\Delta_{1,s}}
    \begin{pmatrix}
    \Delta_{1,1} \\ 
    \Delta_{1,2} \\
    \vdots \\
    \Delta_{1,n}
    \end{pmatrix}\;,  
\end{equation}
which is normalised as $\sum_{j=1}^{n} \rho^{(j)}_{\mathrm{ss}}=1$ with $\rho^{(j)}_{\mathrm{ss}}$ being an element of the density matrix. For $n=2$, the solution~\eqref{eq:steady_n} gives the thermal state of the virtual temperature, $\rho_{\mathrm{ss}}=(\tau_{0,1}^{\mathrm{g}},\tau_{0,1}^{\mathrm{e}})^T$ as expected. 
For $n=3$, the solution~\eqref{eq:steady_n} corresponds to Eq.~\eqref{eq:steady_3}. For the solution~\eqref{eq:steady_n} for $n=4$, see the Appendix \ref{app:steady_4}.

\section{Steady-state solution of effRME for four-level system} \label{app:steady_4}

One can obtain the steady state of the effRME for any-level target system from Eq.~\eqref{eq:steady_n}. In this appendix, we focus on a four-level target system and discuss components of its steady state. Let us recall that the steady state~\eqref{eq:steady_3} of the effRME for qutrit target systems is a combination of other steady states where two pairs of levels are characterised with different temperatures, weighted with the effective thermalisation rates $q_i$. Even for higher-level systems, the same feature can be seen as shown below. 

Suppose that one has a four-level target system where each pair of levels are occupied by one two-qubit machine (in total six machines are involved). The steady-state solution of the effRME is given by 
\begin{align} \label{eq:steady_4}
    \frac{\rho_{\mathrm{ss}}}{C}
    &=
    q_0^3 q_1^3 q_2^3 ~\tau_{012}^{333}
    \nonumber\\
    &
    +q_0^3 q_1^3 q_0^2 ~\tau_{010}^{332}
    +q_0^3 q_1^3 q_1^2 ~\tau_{011}^{332}
    \nonumber\\
    &
    +q_0^3 q_2^3 q_0^1 ~\tau_{020}^{331}
    +q_0^3 q_2^3 q_1^2 ~\tau_{021}^{332}
    \nonumber\\
    &
    +q_1^3 q_2^3 q_0^1 ~\tau_{120}^{331}
    +q_1^3 q_2^3 q_0^2 ~\tau_{120}^{332}
    \nonumber\\
    &
    +q_0^3 q_0^1 q_0^2 ~\tau_{000}^{312}
    +q_0^3 q_0^1 q_1^2 ~\tau_{001}^{312}
    +q_0^3 q_0^2 q_1^2 ~\tau_{001}^{322}
    \nonumber\\
    &
    +q_1^3 q_0^1 q_0^2 ~\tau_{100}^{312}
    +q_1^3 q_0^1 q_1^2 ~\tau_{101}^{312}
    +q_1^3 q_0^2 q_1^2 ~\tau_{101}^{322}
    \nonumber\\
    &
    +q_2^3 q_0^1 q_0^2 ~\tau_{200}^{312}
    +q_2^3 q_0^1 q_1^2 ~\tau_{201}^{312}
    +q_2^3 q_0^2 q_1^2 ~\tau_{201}^{322},
\end{align}
where the normalisation constant $C$ is given by the trace of the right hand side of Eq.~\eqref{eq:steady_4}. 16 states such as $\tau_{012}^{333}$ in the solution are steady states with three of the coherent couplings on. For example, 
\begin{align} \label{eq:tau012333}
   \tau_{012}^{333} 
   &=
   \tau_{0,3}^{\mathrm{g}}\tau_{1,3}^{\mathrm{e}}\tau_{2,3}^{\mathrm{e}} \ketbra{0}{0} 
   + 
   \tau_{0,3}^{\mathrm{e}}\tau_{1,3}^{\mathrm{g}}\tau_{2,3}^{\mathrm{e}} \ketbra{1}{1}
   \nonumber\\
   &+
   \tau_{0,3}^{\mathrm{e}}\tau_{1,3}^{\mathrm{e}}\tau_{2,3}^{\mathrm{g}} \ketbra{2}{2}
   +
   \tau_{0,3}^{\mathrm{e}}\tau_{1,3}^{\mathrm{e}}\tau_{2,3}^{\mathrm{e}} \ketbra{3}{3}
   , 
\end{align}
which is not normalised on purpose, such as Eq.~\eqref{eq:tauABBCCA}. In this state, three pairs of the levels are characterised with different virtual temperatures. 

The steady-state solution~\eqref{eq:steady_4} consists of 16 steady states where three of the coherent couplings are present in the system (e.g. $\tau_{012}^{333}$). However, notice that the solution~\eqref{eq:steady_4} does not cover all the possible steady states with three of the couplings on. 
For example, a case is excluded where three two-qubit machines are coupled to the transitions between $\ket{0}$ and $\ket{1}$, between $\ket{0}$ and $\ket{3}$, and between $\ket{1}$ and $\ket{3}$. 
There are some differences between the excluded cases and the included cases. In the excluded cases, one level is unoccupied. For the above example, the level $\ket{2}$ is free. 
Moreover, the excluded cases are essentially the same as the situation depicted in Fig.~\ref{fig:qutrit_6qubits}, i.e.~the three thermalisation processes compete.
As discussed in Sec.~\ref{sec_qutrit}, this kind of steady states cannot be simply described with just the virtual temperature, but effective rates are required in contrast to Eq.~\eqref{eq:tau012333}. 
In brief, the steady state~\eqref{eq:steady_4} is composed of other steady states that coherently interact with three machines and where the thermalisation processes caused by their machines do not compete against each other. 

\section{Analytical form of the effective thermalisation rates}
\label{app:analyticalqj}

We derive the analytical form of the effective thermalisation rate $q_i$ when we describe the qutrit system depicted in Fig.~{\ref{fig:qutrit_6qubits}} with the GKLSME~{\eqref{eq_master_phys}}. Our interest is to obtain the density matrix $\rho$ of the target qutrit from the GKLSME~{\eqref{eq_master_phys}}. Considering $\Gamma_{i1},\Gamma_{i2}\gg g_i$ for $i\in\{\mathrm{A,B,C}\}$, we use the Nakajima-Zwanzig projection operator technique to reduce the total density matrix to the target qutrit, $\rho=\mathrm{Tr}_{\mathcal{I}}[\rho_{\mathrm{tot}}]$ with $\mathcal{I}\coloneq\{\mathrm{A}1,\mathrm{A}2,\mathrm{B}1,\mathrm{B}2,\mathrm{C}1,\mathrm{C}2\}$. Below, we loosely follow Appendix~D in Ref.~{\cite{Erker2017}}. 

We define a projector $\mathcal{P}$ as
\begin{align}
    \mathcal{P}\rho_{\mathrm{tot}}(t)
    &=
    \rho(t)\otimes\tau_{\mathrm{A}1}\otimes\tau_{\mathrm{A}2}\otimes\tau_{\mathrm{B}1}\otimes\tau_{\mathrm{B}2}\otimes\tau_{\mathrm{C}1}\otimes\tau_{\mathrm{C}2}.
\end{align}
Rewriting Eq.~{\eqref{eq_master_phys}} as $\partial\rho_{\mathrm{tot}}/dt=\mathcal{L}\rho_{\mathrm{tot}}$, we decompose the Liouvillian as $\mathcal{L}=\mathcal{L}_0+\mathcal{H}_{\mathrm{int}}$, where the Hamiltonian superoperator $\mathcal{H}_{\mathrm{int}}$ is defined as
\begin{align}
    \mathcal{H}_{\mathrm{int}}\rho_{\mathrm{tot}}
    &=
    i\left(
    \rho_{\mathrm{tot}}H_{\mathrm{int}}^{\dagger}
    -
    H_{\mathrm{int}}\rho_{\mathrm{tot}}
    \right),
\end{align}
where $H_{\mathrm{int}}=g_\mathrm{A}\ketbra{0}{1}\sigma^+_{\mathrm{A}1}\sigma^-_{\mathrm{A}2}+g_\mathrm{B}\ketbra{0}{2}\sigma^+_{\mathrm{B}1}\sigma^-_{\mathrm{B}2}+g_\mathrm{C}\ketbra{1}{2}\sigma^+_{\mathrm{C}1}\sigma^-_{\mathrm{C}2}+\mathrm{H.c.}$ as seen in Eq.~{\eqref{eq_H}}.
We adapt a dissipative interaction picture and transform the total density matrix as $\tilde{\rho}_{\mathrm{tot}}(t)=\mathrm{e}^{-\mathcal{L}_0 t}\rho_{\mathrm{tot}}(t)$ and the superoperator as $\tilde{\mathcal{H}}_{\mathrm{int}}(t)=\mathrm{e}^{-\mathcal{L}_0 t}\mathcal{H}_{\mathrm{int}}\mathrm{e}^{\mathcal{L}_0 t}$. After going through the standard perturbative argument~{\cite{breuerbook}}, we have
\begin{align} \label{eq:der_t_rho}
    \frac{d\mathcal{P}\tilde{\rho}_{\mathrm{tot}}}{dt}
    &=
    \int_0^t\!\! dt' \,\,
    \mathcal{P}
    \tilde{\mathcal{H}}_{\mathrm{int}}(t)
    \tilde{\mathcal{H}}_{\mathrm{int}}(t')
    \mathcal{P}
    \tilde{\rho}_{\mathrm{tot}}(t'),
\end{align}
which is valid to the second order of $g_i/\Gamma_{i1}$ and $g_i/\Gamma_{i2}$. Here, we take several steps. Due to $\Gamma_{i1},\Gamma_{i2}\gg g_{i}$, we use the Born-Markov approximation to the integral, expand the lower integration limit to $-\infty$, and replace $\tilde{\rho}_{\mathrm{tot}}(t')\to\tilde{\rho}_{\mathrm{tot}}(t)$. To simplify Eq.~{\eqref{eq:der_t_rho}}, we expand the commutators, trace out the machines, and transform back to the Schr\"{o}dinger picture. In the end, the master equation describes the time evolution of the diagonal terms (population) and the off-diagonal terms (coherence) independently. Here, we focus on the population, and its equation is given by
\begin{align}
    \frac{d \rho}{d t}
    &=
    \sum_{i=\mathrm{A},\mathrm{B},\mathrm{C}}
    p_{i}^{\downarrow}
    \mathcal{D}[O_i]\rho
    +
    p_{i}^{\uparrow}
    \mathcal{D}[O_{i}^{\dagger}]\rho
    ,
\end{align}
where $O_{\mathrm{A}}=\ketbra{0}{1}$, $O_{\mathrm{B}}=\ketbra{0}{2}$, $O_{\mathrm{C}}=\ketbra{1}{2}$, and $p_{i}^{\downarrow}$ and $p_{i}^{\uparrow}$ are the backward and forward rates, respectively. Defining a vector of the diagonal terms in the qutrit density matrix as $\vec{\rho}_{\mathrm{diag}}=(\rho^{(0)},\rho^{(1)},\ldots,\rho^{(n-1)})^{T}$, the above equation becomes
\begin{align} \label{eq:equation_population}
    \frac{d \vec{\rho}_{\mathrm{diag}}}{dt}
    &=
    \bold{M}_1
    \vec{\rho}_{\mathrm{diag}}
    ,
\end{align}
where 
\begin{align} \label{eq:matrixM1}
    \bold{M}_1
    &=
    \begin{pmatrix}
    -p_{\mathrm{A}}^{\uparrow}-p_{\mathrm{B}}^{\uparrow} & p_{\mathrm{A}}^{\downarrow} & p_{\mathrm{B}}^{\downarrow} \\
    p_{\mathrm{A}}^{\uparrow} & -p_{\mathrm{C}}^{\uparrow}-p_{\mathrm{A}}^{\downarrow} & p_{\mathrm{C}}^{\downarrow} \\
    p_{\mathrm{B}}^{\uparrow} & p_{\mathrm{C}}^{\uparrow} & -p_{\mathrm{B}}^{\downarrow}-p_{\mathrm{C}}^{\downarrow}
    \end{pmatrix}\;.
\end{align}
The backward and forward rates $p_{i}^{\downarrow}$, $p_{i}^{\uparrow}$ are given by
\begin{align}
    p_{i}^{\downarrow}
    &=
    2g_i^2
    \int_0^{\infty}\!\! dt \,\, \mathrm{e}^{i\omega_i t}
    \langle
    \sigma_{i1}(t)\sigma_{i2}^{\dagger}(t)
    \sigma_{i1}^{\dagger}(0)\sigma_{i2}(0)
    \rangle_i
    \nonumber\\
    &=
    \frac{
    4g_i^2\tau_{i1}^{\mathrm{g}}\tau_{i2}^{\mathrm{e}}}{
    \Gamma_{i1}\left(\bar{n}(\Omega_{i1},T_{i1})+1\right)\mathcal{Z}_{i1}
    +
    \Gamma_{i2}\left(\bar{n}(\Omega_{i2},T_{i2})+1\right)\mathcal{Z}_{i2}
    }
\end{align}
and
\begin{align}
    p_{i}^{\uparrow}
    &=
    2g_i^2
    \int_0^{\infty}\!\! dt \,\, \mathrm{e}^{-i\omega_i t}
    \langle
    \sigma_{i1}^{\dagger}(t)\sigma_{i2}(t)
    \sigma_{i1}(0)\sigma_{i2}^{\dagger}(0)
    \rangle_i
    \nonumber\\
    &=
    \frac{
    4g_i^2\tau_{i1}^{\mathrm{e}}\tau_{i2}^{\mathrm{g}}}{
    \Gamma_{i1}\left(\bar{n}(\Omega_{i1},T_{i1})+1\right)\mathcal{Z}_{i1}
    +
    \Gamma_{i2}\left(\bar{n}(\Omega_{i2},T_{i2})+1\right)\mathcal{Z}_{i2}
    }
    ,
\end{align}
where $\omega_{\mathrm{A}}=\omega_1-\omega_0$, $\omega_{\mathrm{B}}=\omega_2-\omega_0$, $\omega_{\mathrm{C}}=\omega_2-\omega_1$, $\sigma_i(t)=\mathrm{e}^{\mathcal{L}_0^{\dagger}t}\sigma_i$, $\langle\cdots\rangle_i=\mathrm{Tr}[\tau_{i1}\otimes\tau_{i2}\cdots]$, and $\mathcal{Z}_{i1,2}=1+\mathrm{e}^{-\beta_{i1,2}\omega_{i}}$ are the partition functions. 
The adjoint Liouvillian $\mathcal{L}_0^{\dagger}$ is defined as $\mathrm{Tr}[Q\mathcal{L}_0(P)]=\mathrm{Tr}[\mathcal{L}_0^{\dagger}(Q)P]$ for any operators $P$ and $Q$. 
Notice that $p_{i}^{\uparrow}/p_{i}^{\downarrow}=\mathrm{e}^{-\omega_i/T_{\mathrm{v}i}}$. This model can be regarded as a biased random walk~{\cite{Erker2017}}. 

It is important to note that if we consider the steady state regime and so $d\vec{\rho}_{\mathrm{diag}}/dt=0$, Eq.~{\eqref{eq:equation_population}} corresponds to Eq.~{\eqref{eq:Mrho0}}, and the matrix~{\eqref{eq:matrixM1}} can agree with the matrix~{\eqref{eq:matrixM0}} after multiplying with a constant: $\alpha\bold{M}_1=\bold{M}$. This means that Eq.~{\eqref{eq:equation_population}} can be written in the same form of the effRME~{\eqref{eq:master_n}} when looking at the population, and in this case we have
\begin{align} \label{eq:effectiveq_GKLSME}
    q_{i}
    &=
    \alpha\left(
    p_{i}^{\downarrow} + p_{i}^{\uparrow}
    \right)
    \nonumber\\
    &=
    \frac{
    4\alpha g_i^2
    \left(
    \tau_{i1}^{\mathrm{g}}\tau_{i2}^{\mathrm{e}}
    +
    \tau_{i1}^{\mathrm{e}}\tau_{i2}^{\mathrm{g}}
    \right)
    }{
    \Gamma_{i1}\left(\bar{n}(\Omega_{i1},T_{i1})+1\right)\mathcal{Z}_{i1}
    +
    \Gamma_{i2}\left(\bar{n}(\Omega_{i2},T_{i2})+1\right)\mathcal{Z}_{i2}
    }
\end{align}
for $i=\mathrm{A,B,C}$. In the main text, we refer to this form in Eq.~{\eqref{eq:effectiveq_GKLSME_text}} and take $\alpha=1/2$ as explained at the end of this appendix. As expected, $q_{i}$ has $g_i^2$-dependence, and its temperature dependence corresponds to the norm of the virtual qubit multiplied with an additional term. The constant $\alpha$ does not really matter since the ratios $q_i/q_j$ of the effective rates appear in the steady state rather than $q_i$ itself. This derivation can be easily extended to $n$-level target systems considered in Appendix~{\ref{app:derivation_steady_n}}.

As shown in Ref.~{\cite{rignon2021thermodynamics}}, the RME~{\eqref{eq:master_reset}} we consider in Sec.~{\ref{sec:RME}} can be mapped to a GKLSME for $Q_{i1}=Q_{i2}=Q_i$, such as
\begin{align}
    \frac{\partial \rho_{\mathrm{tot}}}{\partial t}
    =
    -i\left[H,\rho_{\mathrm{tot}}\right]
    +
    \sum_{j\in\mathcal{J}}
    \big(
    &
    Q_{j}^{+}\mathcal{D}[\sigma_j^{+}] +
    Q_{j}^{-}\mathcal{D}[\sigma_j^{-}] 
    \nonumber\\
    &
    + 
    Q^{z}\mathcal{D}[\sigma_j^{z}]
    \big)\rho_{\mathrm{tot}},
\end{align}
where $Q_{j}^{+}=Q_{j}\mathrm{e}^{-\beta_j \Omega_j}/\mathcal{Z}_j$, $Q_{j}^{-}=Q_{j}/\mathcal{Z}_j$, $Q_{j}^{z}=Q_{j}/4$, and $\mathcal{J}\coloneq\{\mathrm{A}1,\mathrm{A}2,\mathrm{B}1,\mathrm{B}2,\mathrm{C}1,\mathrm{C}2\}$.
This GKLSME has an additional local dephasing term compared to the GKLSME~{\eqref{eq_master_phys}}. This local dephasing term does not affect the population distribution, and thus, one can obtain the effective rates with the same technique shown above. As a result, the effective rates in the RME~{\eqref{eq:master_reset}} are given by
\begin{align} \label{eq:effectiveq_RME}
    q_i
    &=
    \frac{
    2\alpha g_i^2
    \left(
    \tau_{i1}^{\mathrm{g}}\tau_{i2}^{\mathrm{e}}
    +
    \tau_{i1}^{\mathrm{e}}\tau_{i2}^{\mathrm{g}}
    \right)
    }{
    Q_{i}
    }.
\end{align}
We take $\alpha=1/2$ to make Eq.~{\eqref{eq:effectiveq_RME}} consistent with the effective rate~{\eqref{eq:qvir}} of the qubit target system.

\end{document}